\documentclass[aps, prb, reprint, superscriptaddress]{revtex4-1}
\usepackage{amssymb}
\usepackage{amsmath}
\usepackage{graphicx}
\usepackage{bm}
\usepackage{color}
\usepackage{lineno}

\begin{document}

\title{Efficient heat-energy conversion from a non-thermal
Tomonaga-Luttinger liquid}
\author{Hikaru Yamazaki}
\thanks{These authors contributed equally}
\author{Masashi Uemura}
\thanks{These authors contributed equally}
\author{Haruhi Tanaka}
\author{Tokuro Hata}
\author{Chaojing Lin}
\affiliation{Department of Physics, Institute of Science Tokyo, 2-12-1 Ookayama, Meguro,
Tokyo, 152-8551, Japan.}
\author{Takafumi Akiho}
\author{Koji Muraki}
\affiliation{Basic Research Laboratories, NTT Inc., 3-1 Morinosato-Wakamiya, Atsugi
243-0198, Japan.}
\author{Toshimasa Fujisawa}
\email{fujisawa@phys.sci.isct.ac.jp}
\affiliation{Department of Physics, Institute of Science Tokyo, 2-12-1 Ookayama, Meguro,
Tokyo, 152-8551, Japan.}
\date{\today }
\maketitle

\textbf{Energy harvesting is a technique that generates useful work from
waste heat. Conventional energy harvesters acting on local thermal
equilibrium states are constrained by thermodynamic limits, such as the
Carnot efficiency. Quantum heat engines with non-thermal reservoirs are
expected to exceed such limits. Here, we demonstrate energy harvesting from
a non-thermal Tomonaga-Luttinger (TL) liquid in quantum Hall edge channels,
where the non-thermal state is naturally formed due to the absence of
thermalization. The scheme is tested with a quantum-dot energy harvester
working on a non-thermal TL liquid supplied with waste heat from a
quantum-point-contact transistor. Compared to the quasi-thermalized TL
liquid, the non-thermal state prepared under the same heat is capable of a
larger electromotive force and higher conversion efficiency. These
characteristics can be understood by considering a binary Fermi distribution
function of the non-thermal state induced by entropy-conserving
equilibration. TL liquids are attractive non-thermal carriers for excellent
energy harvesting.}

\bigskip

{\large Introduction}

Quantum heat engines, in which the work is generated by quantum mechanical
processes, are attractive for outstanding performances \cite%
{MyersAVS2022,BookQuantumThermodynamics}. Particularly, non-thermal states
are recognized as useful energy sources to produce higher output power, as
compared to thermalized states. Examples include a specific coherent state
in a three-level system \cite{ScullyScience2003}, squeezed thermal reservoir 
\cite{RossnagelPRL2014,KlaersPRX2017}, coherent atom ensemble showing
superradiance \cite{KimNatPhoto2022}, nonequilibrium cold atoms prepared by
laser cooling \cite{MayerCommPhys2023}, and so on. These non-thermal states
are intentionally produced with specific techniques, which may not be
convenient for energy-harvesting applications. Instead, one can consider an
integrable system as a working fluid, where the system never relaxes to
thermalized states \cite{KennesPRB2017,ChenNPJQI2019}. Even in the presence
of small non-integrable perturbation, long-lived non-thermal states
(prethermalized states) can appear \cite%
{KinoshitaNature2006,BlochRMP2008,KollarPRB2011}. Non-thermal states can be
generated simply by supplying waste heat to an integrable system, and the
non-thermal heat can be recycled into useful work by an energy harvester.
Among experimentally accessible integrable systems, Tomonaga-Luttinger (TL)
liquids in quantum Hall edge channels are attractive for this purpose \cite%
{BookGiamarchi}. First, the non-thermal states prepared with a heat source
can be investigated using an energy spectrometer \cite%
{Altimiras-NatPhys10,leSueurPRL2010,WashioPRB2016,Itoh-PRL2018}. Second,
unidirectional heat transport with quantized heat conductance efficiently
delivers heat to the engine with minimal energy loss \cite%
{JezouinScience2013,SivreNatComm2019,KonumaPRB2022}. Such heat transport has
been discussed with heat Coulomb blockade, anyonic heat flow, and so on \cite%
{SivreNatPhys2018,RosenblattNatComm2017,RouraBasPRB2018}. However, the
advantage of non-thermal TL liquids on thermoelectricity has not been
elucidated.

Here, we propose and demonstrate a prototypical energy harvesting scheme
with a non-thermal TL liquid, as shown in Fig. 1a. Waste heat generated from
an active device is drawn into the TL liquid, and a fraction of the waste
heat is converted into electrical power using a heat engine. We shall show
below that the performance for a non-thermal (NT) state is superior, as
compared to that for an experimentally available quasi-thermalized (QT)
state. The NT state provides a larger electromotive force and higher maximum
conversion efficiency in the zero power limit. Higher conversion efficiency
at maximum power generation is attractive for energy harvesting
applications. A strategy for a higher energy recovery ratio is also shown.
The results encourage us to utilize a TL liquid as a non-thermal energy
resource.

\begin{figure*}[tb]
\begin{center}
\includegraphics[width = 7in]{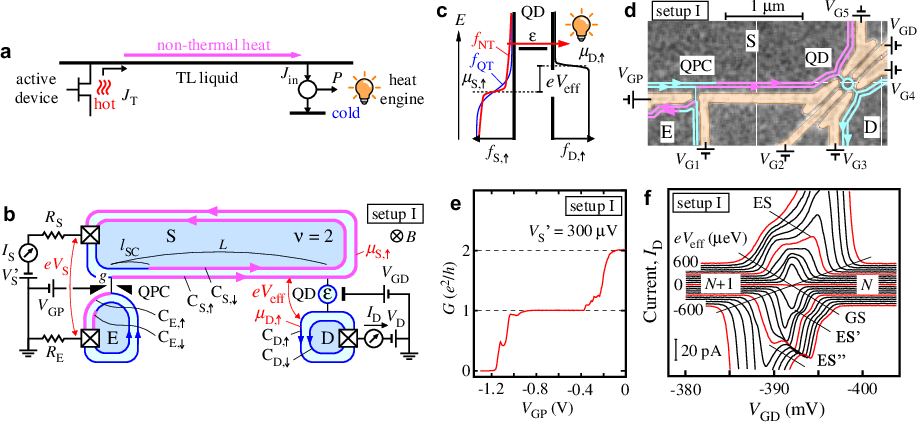}
\end{center}
\caption{ \textbf{Setup for energy-harvesting experiment. }\textbf{a},
Schematic heat flow from an active device through a Tomonaga-Luttinger
(TL) liquid to a heat engine. The heat power $J_{\mathrm{T}}$\
generated from the device is converted to electric power $P$\ through a
non-thermal TL liquid. \textbf{b}, Schematic device structure with emitter
(E), source (S), and drain (D) regions. Electrons and heat travel
unidirectionally in the edge channels C$_{r,\uparrow }$ and C$_{r,\downarrow
}$ of $r\in \left\{ \mathrm{E,S,D}\right\} $. The quantum point contact
(QPC) with bias voltage $V_{\mathrm{S}}$ and transmission coefficient $g$
generates total heat current $J_{\mathrm{T}}$ in the channels (magenta
lines). The quantum dot (QD) with an effective bias voltage, $V_{\mathrm{%
eff}}$, works as a heat engine. Thermoelectricity for non-thermal (NT) and
quasi-thermal (QT) states is evaluated by measuring the drain current $%
I_{\mathrm{D}}$. \textbf{c}, Schematic energy diagram of the QD heat engine
for NT and QT distribution functions $f_{\mathrm{NT}}$ and $f_{\mathrm{QT}}$%
, respectively, in the source. As compared to the QT state, the NT state
under the same $J_{\mathrm{T}}$\ provides higher electromotive force and
higher thermoelectric efficiency. \textbf{d}, Scanning electron micrograph
of a control device with false color for setup I. The distance between the
QPC and the QD is $L=$ 2 $\protect\mu $m. \textbf{e}, Conductance $G$ of
the QPC. The NT and QT states are prepared at $G\sim $\ 0.03 $e^{2}/h$\ and $%
\sim $\ 0.5 $e^{2}/h$, respectively. \textbf{f}, Coulomb diamond
characteristics of the QD. The QD current $I_{\mathrm{D}}$\ is plotted as a
function of the gate voltage $V_{\mathrm{GD}}$\ for various effective bias
voltage $V_{\mathrm{eff}}$. Each trace is offset for clarity. Current
steps are associated with transport through the ground state (GS) and
excited states (ES, ES', and ES\textquotedblright ) of the QD.}
\end{figure*}

\bigskip

{\large Results and Discussion}

\textbf{Integrated heat circuit.} We employ an integrated heat circuit shown
in Fig. 1b consisting of a quantum point contact (QPC) as an active device
and a quantum dot (QD) as a heat engine in the quantum Hall regime at
Landau-level filling factor $\nu $ = 2. Two chiral edge channels for spin-up
and -down, labeled C$_{r,\uparrow }$ and C$_{r,\downarrow }$, respectively,
are formed along the edge of each quantum Hall region ($r=$ E for emitter, S
for source, and D for drain) \cite{KonumaPRB2022}. The interaction between
the copropagating channels constitutes a TL liquid on each edge. The
channels emanating from ohmic contacts (crossed boxes) are in thermal
equilibrium at base electron temperature $T_{\mathrm{B}}$, as represented by
the blue lines. The QPC conductance $G=ge^{2}/h$ is tuned by the gate
voltage $V_{\mathrm{GP}}$. In the tunneling regime at $0<g<1$ relevant to
the present experiment, the total heat power $J_{\mathrm{T}}=\frac{e^{2}}{h}%
g(1-g)V_{\mathrm{S}}^{2}$ for the effective bias voltage $V_{\mathrm{S}}$ is
distributed firstly into the channels C$_{\mathrm{E},\uparrow }$ and C$_{%
\mathrm{S},\uparrow }$ by the stochastic partition process, and then
fractionalized into the four channels C$_{\mathrm{E},\uparrow }$, C$_{%
\mathrm{E},\downarrow }$, C$_{\mathrm{S},\uparrow }$, and C$_{\mathrm{S}%
,\downarrow }$, represented by the thick magenta lines, due to the
spin-charge separation characteristic of TL liquids \cite%
{FreulonNatComm-SC,Bocquillon-NatCom2013,Inoue-PRL2014,Hashisaka-NatPhys2017}%
. As the spin-charge separation is deterministic, the electronic states in
the channels remain non-thermal \cite%
{GutmanPRL2008,Iucci2009,GutmanPRB2010,KovrizhinPRB2011,LevkivskyiPRB2012}.

By choosing small $g\ll \frac{1}{2}$, the energy distribution function $f_{%
\mathrm{S,\uparrow }}(E)$ in the source channel C$_{\mathrm{S},\uparrow }$
can deviate significantly from the thermalized Fermi distribution function $%
f_{\mathrm{th}}(E)$. This study aims to evaluate the NT state, prepared in
this manner and described by the distribution function $f_{\mathrm{NT}}(E)$,
as a heat source. Since a thermalized state is unavailable, we use a QT
state prepared at $g\simeq \frac{1}{2}$ as a reference, which has a
distribution function $f_{\mathrm{QT}}(E)$ close to $f_{\mathrm{th}}(E)$.
Namely, we compare the performances for the NT and QT states generated by
the same $J_{\mathrm{T}}$ by tuning $g$ and $V_{\mathrm{S}}$. The NT state
contains an excess of high-energy electrons well above the chemical
potential $\mu $ and low-energy holes well below $\mu $, as schematically
shown in Fig. 1c with $f_{\mathrm{NT}}(E)$, while the heat current remains
identical to that for $f_{\mathrm{QT}}(E)$. It should be noted that this NT
state is not a trivial non-thermal state that can be prepared, for example,
by mixing two thermal baths at a point contact \cite{SanchezPRL2019}. The NT
state in the TL liquid is a stationary and equilibrated state that
theoretically never relaxes to a thermalized state. This comes from the
integrable model of the TL liquids\cite%
{BookGiamarchi,WashioPRB2016,Itoh-PRL2018} and is attractive for energy
harvesting.

We use a QD heat engine to evaluate the NT and QT states as a heat source. 
\cite{EspositoEuroPhysLett2009,JosefssonNatNano2018}. For example, suppose
that electrons at energy $\varepsilon $ in the source (C$_{\mathrm{S}%
,\uparrow }$) with chemical potential $\mu _{\mathrm{S},\uparrow }$ are
transferred to the drain (C$_{\mathrm{D},\uparrow }$) with chemical
potential $\mu _{\mathrm{D},\uparrow }$ ($>\mu _{\mathrm{S},\uparrow }$).
This transport induces a negative current $I_{\mathrm{D}}$ ($<0$) against a
positive effective voltage $V_{\mathrm{eff}}=\left( \mu _{\mathrm{D}%
,\uparrow }-\mu _{\mathrm{S},\uparrow }\right) /e$ ($>0$) in the setup of
Fig. 1b to generate net electric power $P=-I_{\mathrm{D}}V_{\mathrm{eff}}$ ($%
>0$). The high-energy electrons in the NT state are expected to provide
large electromotive force $V_{\mathrm{emf}}$ defined as maximum $V_{\mathrm{%
eff}}$ to produce finite $P>0$.

Significantly, the heat conversion efficiency for the NT state can exceed
that for the thermalized or QT state. For simplicity, we consider an
idealized QD with a single energy level $\varepsilon $, described by a
delta-function-type energy transmission function, neglecting lifetime
broadening and excited states\cite{HumphreyPRL2002}. As single-electron
transport extracts heat energy $\varepsilon -\mu _{\mathrm{S},\uparrow }$
from the source to produce electric energy $eV_{\mathrm{eff}}=\mu _{\mathrm{D%
},\uparrow }-\mu _{\mathrm{S},\uparrow }$ while discarding the remaining
heat $\varepsilon -\mu _{\mathrm{D},\uparrow }$ to the drain, the conversion
efficiency of the idealized engine is given by $\bar{\eta}=eV_{\mathrm{eff}%
}/\left( \varepsilon -\mu _{\mathrm{S},\uparrow }\right) $. We evaluate this 
$\bar{\eta}$ for the NT and QT states in the following experiments.

One can evaluate efficiency for a realistic QD. However, this is not
convenient for comparing NT and QT states experimentally because the
characteristics depend on the detail (level broadening and excited states)
of the QD in different ways for NT and QT states. We believe the idealized
efficiency $\bar{\eta}$ is suitable for evaluating the non-thermal heat
source. Nevertheless, we use a realistic QD to reveal the power-generation
conditions ($eV_{\mathrm{eff}}$, $\varepsilon $, and $\mu _{\mathrm{S}%
,\uparrow }$) for each state, from which we estimate $\bar{\eta}$.

\textbf{Measurements.} We experimentally demonstrate the energy harvesting
scheme using a device fabricated in a standard AlGaAs/GaAs heterostructure
with an electron density of 3.1$\times $10$^{11}$ cm$^{-2}$ (see
Supplementary Fig. 1 for the gate pattern). We present data from two
measurement setups (setup I in Fig. 1b and setup II in Supplementary Fig.
2a), in which the QD exhibited different characteristics. All measurements
were performed at magnetic field $B=$ 6 T ($\nu =$ 2) and $T_{\mathrm{B}%
}\simeq $ 150 mK ($k_{\mathrm{B}}T_{\mathrm{B}}\simeq $ 13 $\mu $eV). As
shown in Fig. 1d for setup I, we formed a QPC transistor and a QD heat
engine with appropriate gate voltages on the gates (yellow regions). The QPC
conductance $G=I_{\mathrm{S}}/V_{\mathrm{S}}$ in Fig. 1e (Supplementary Fig.
2c for setup II) shows clear quantized conductances $G=ge^{2}/h$ at $g=$ 1
and 2. Here, we choose the tunneling regime at $g\simeq 0.03$ and $0.5$ to
prepare NT and QT states, respectively. We adjusted the supply voltage $V_{%
\mathrm{S}}^{\prime }$ under the series resistance $R_{\mathrm{S}}+R_{%
\mathrm{E}}$ to maintain the same generated heat power $J_{\mathrm{T}}$ for
both. The applied voltage $\left\vert V_{\mathrm{S}}\right\vert =$ 30 - 800 $%
\mu $V is sufficiently smaller than the upper limit (a few mV) where the
characteristic non-thermal excitation remains \cite{SuzukiCommPhys2023}.

The QD is attached to the source channel with tunnel rate $\Gamma _{\mathrm{S%
}}$ and drain channel with tunnel rate $\Gamma _{\mathrm{D}}$, which are
characterized by the overall tunnel rate $\Gamma =\Gamma _{\mathrm{S}}\Gamma
_{\mathrm{D}}/\left( \Gamma _{\mathrm{S}}+\Gamma _{\mathrm{D}}\right) $ and
the resonant width $w=\hbar \left( \Gamma _{\mathrm{S}}+\Gamma _{\mathrm{D}%
}\right) /2$. Figure 1f shows the Coulomb diamond characteristics of the QD
in setup I with $\Gamma _{\mathrm{S}}\simeq $ 0.16 GHz and $\Gamma _{\mathrm{%
D}}\simeq $ 5.7 GHz ($\Gamma \simeq $\ 0.2 GHz and $w\simeq $\ 2 $\mu $eV)
under no heat injection ($J_{\mathrm{T}}=0$ and $g=0$). The current steps
associated with the ground state (GS) and excited states (ES, ES$^{\prime }$%
, and ES$^{\prime \prime }$) of $N$- and $\left( N+1\right) $-electron QD
indicate level spacing $\Delta _{1}\simeq $ 400 $\mu $eV between GS and ES
of $N$-electron QD, $\Delta _{1}^{\prime }\simeq $ 80 $\mu $eV between GS
and ES$^{\prime }$, and $\Delta _{2}^{\prime }\simeq $ 400 $\mu $eV between
GS and ES$^{\prime \prime }$ of $\left( N+1\right) $-electron QD. The QD in
setup II shows $\Gamma _{\mathrm{S}}\simeq \Gamma _{\mathrm{D}}\simeq $ 1
GHz ($\Gamma \simeq $\ 0.5 GHz and $w=$ 0.6 $\mu $eV) and $\Delta _{1}\simeq
\Delta _{1}^{\prime }\simeq $ 400 $\mu $eV (see Supplementary Fig. 2d). For
both setups, $w$ is sufficiently narrow ($<k_{\mathrm{B}}T_{\mathrm{B}%
}\simeq $ 13 $\mu $eV), and thus the level broadening can be neglected for
simplicity. The excited states in our QDs can be neglected in the evaluation
of $V_{\mathrm{emf}}$ and $\bar{\eta}$ but modify $P$, as we discuss later.

The distance $L=$ 2 $\mu $m (2.1 $\mu $m in setup II) between the QD and the
QPC is sufficiently long compared to the length $l_{\mathrm{SC}}$ ($\lesssim 
$ 0.1 $\mu $m) required for the electronic system to reach the non-thermal
steady state, meaning that the electron-electron interaction is
significantly strong. The system would be thermalized, if it were a
conventional Fermi liquid\cite{PothierPRL1997}. The distance is well below
the dissipation length of about 20 $\mu $m to the environment (other than
the two copropagating channels), where $J_{\mathrm{T}}$ is distributed in
the four channels almost equally\cite{Itoh-PRL2018}.

\begin{figure*}[tb]
\begin{center}
\includegraphics[width = 7in]{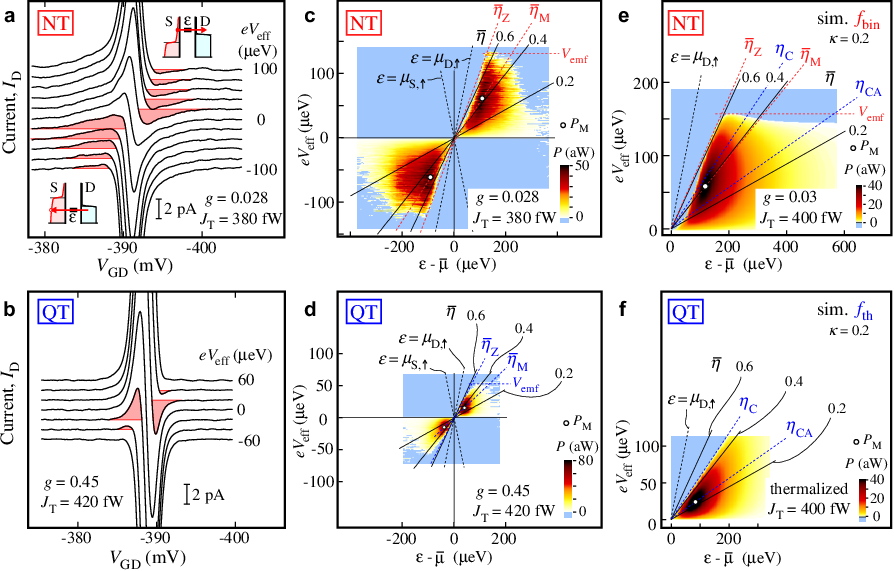}
\end{center}
\caption{\textbf{Heat-engine characteristics for non-thermal (NT) and
quasi-thermal (QT) states.} \textbf{a}, \textbf{b}, Drain\ current $I_{%
\mathrm{D}}$ as a function of the gate voltage $V_{\mathrm{GD}}$ at various
effective energy bias $eV_{\mathrm{eff}}$. Positive power generation is
highlighted by red. The upper-right and lower-left insets in \textbf{a} show
the energy diagrams for positive power generation at $eV_{\mathrm{eff}}>0$ ($%
\protect\varepsilon >\protect\mu _{\mathrm{D},\uparrow }>\protect\mu _{%
\mathrm{S},\uparrow }$) and $eV_{\mathrm{eff}}<0$ ($\protect\varepsilon <%
\protect\mu _{\mathrm{D},\uparrow }<\protect\mu _{\mathrm{S},\uparrow }$),
respectively, where $\protect\varepsilon $ is the electrochemical
potential of the quantum dot (QD), and $\protect\mu _{\mathrm{D},\uparrow }$
and $\protect\mu _{\mathrm{S},\uparrow }$ are the chemical potentials of the
drain and source spin-up channels, respectively. \textbf{c}, \textbf{d},
Color plot of electric power $P=-I_{\mathrm{D}}V_{\mathrm{eff}}$ as a
function of $eV_{\mathrm{eff}}$ and $\protect\varepsilon -\bar{\protect\mu}$%
, where $\bar{\protect\mu}=\left( \protect\mu _{\mathrm{D},\uparrow }+%
\protect\mu _{\mathrm{S},\uparrow }\right) /2$ is the average chemical
potential. Comparable heat power $J_{\mathrm{T}}\simeq $ 400 fW was used
in \textbf{a} and \textbf{c} for the NT state, and \textbf{b} and \textbf{d}
for the QT state. The NT state provides high electromotive force $V_{%
\mathrm{emf}}\simeq $\ 130 $\protect\mu $V and high efficiencies $\bar{%
\protect\eta}_{\mathrm{Z}}\simeq $\ 0.65 in the zero power limit and $%
\bar{\protect\eta}_{\mathrm{M}}\simeq $\ 0.45 at maximum power in c, as
compared to the QT state ($V_{\mathrm{emf}}\simeq $\ 50 $\protect\mu $V, $%
\bar{\protect\eta}_{\mathrm{Z}}\simeq $\ 0.58 and $\bar{\protect\eta}_{%
\mathrm{M}}\simeq $\ 0.35) in d. \textbf{e}, \textbf{f}, Color plot of $P$
calculated for an idealized QD by using the binary Fermi distribution
function $f_{\mathrm{bin}}$ with the fraction $p=$ 0.12, the high
thermal energy $k_{\mathrm{B}}T_{\mathrm{S}}=$ 116 $\protect\mu $eV, the
low thermal energy $k_{\mathrm{B}}T_{\mathrm{L}}=$ 15.2 $\protect\mu $eV,
and the heat transfer factor $\protect\kappa =$ 0.2\ for the NT state in 
\textbf{e} and the thermalized Fermi distribution function $f_{\mathrm{th%
}}$ with the thermal energies $k_{\mathrm{B}}T_{\mathrm{th}}=$ 42.3 $%
\protect\mu $eV in the source and $k_{\mathrm{B}}T_{\mathrm{th}}^{\prime
}=$ 23.2 $\protect\mu $eV in the drain for the thermalized state in 
\textbf{f}. The conditions for $\protect\varepsilon =\protect\mu _{\mathrm{D}%
,\uparrow }$ and $\protect\varepsilon =\protect\mu _{\mathrm{S},\uparrow }$,
and the idealized efficiency $\bar{\protect\eta}=$ 0.2, 0.4, and 0.6 are
shown by the dashed and solid lines, respectively, in \textbf{c }- \textbf{f}%
. The estimated $\bar{\protect\eta}_{\mathrm{Z}}$ and $\bar{\protect\eta}_{%
\mathrm{M}}$ are shown by red and blue dashed lines in \textbf{c}-\textbf{f}%
, while those in \textbf{f} coincide with the Carnot efficiency $\protect%
\eta _{\mathrm{C}}$ and the Curzon-Ahlborn efficiency $\protect\eta _{%
\mathrm{CA}}$, respectively.}
\end{figure*}

As an example of energy harvesting experiments, we plot the drain current $%
I_{\mathrm{D}}$ as a function of the QD gate voltage $V_{\mathrm{GD}}$ in
Fig. 2a for an NT state ($g=0.028$, $V_{\mathrm{S}}=600$ $\mu $V) and in
Fig. 2b for a QT state ($g=0.45$, $V_{\mathrm{S}}=210$ $\mu $V), both
obtained under comparable $J_{\mathrm{T}}\simeq $ 400 fW. Power generation
with $P>0$ ($I_{\mathrm{D}}<0$ at $V_{\mathrm{eff}}>0$ and $I_{\mathrm{D}}>0$
at $V_{\mathrm{eff}}<0$, highlighted by red) is seen in the vicinity of the
Coulomb blockade peak. Power generation\ is observed across the wide range
of $\varepsilon $ and $eV_{\mathrm{eff}}$ for the NT state. While power
generation ceases at $\left\vert V_{\mathrm{eff}}\right\vert \gtrsim $ 60 $%
\mu $V for the QT state, it persists up to a much larger $\left\vert V_{%
\mathrm{eff}}\right\vert >$ 100 $\mu $V for the NT state.

\begin{figure*}[tb]
\begin{center}
\includegraphics[width = 6in]{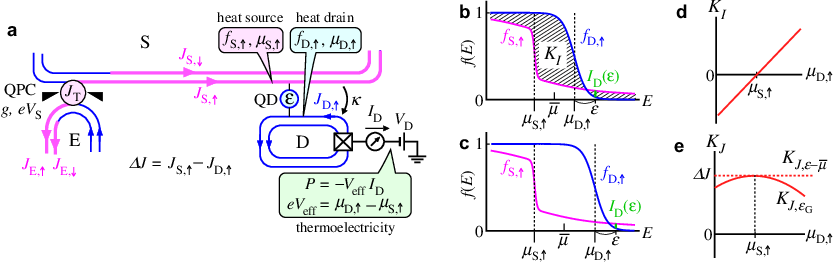}
\end{center}
\caption{\textbf{Analysis of heat-engine characteristics.} \textbf{a},
Schematic diagram of the analysis. The total heat power $J_{\mathrm{T}}$
generated at the quantum point contact (QPC) is divided into non-thermal
heat currents $J_{\mathrm{S,\uparrow }}$, $J_{\mathrm{S,\downarrow }}$, $J_{%
\mathrm{E,\uparrow }}$, and $J_{\mathrm{E,\downarrow }}$. The long-range
interaction with heat transfer factor $\protect\kappa $ induces heat current 
$J_{\mathrm{D,\uparrow }}$ in the drain channel. The quantum dot (QD)
heat engine is attached to the heat source with a distribution function $%
f_{\mathrm{S,\uparrow }}$ and a chemical potential $\protect\mu _{\mathrm{%
S,\uparrow }}$, and the heat drain with $f_{\mathrm{D,\uparrow }}$ and $%
\protect\mu _{\mathrm{D,\uparrow }}$. The net power $P$ generated by the QD
is estimated from the effective bias $V_{\mathrm{eff}}$ and the measured
current $I_{\mathrm{D}}$. \textbf{b}, \textbf{c}, Distribution functions $f_{%
\mathrm{S,\uparrow }}$ and $f_{\mathrm{D,\uparrow }}$. The current $I_{%
\mathrm{D}}=-e\Gamma \left[ f_{\mathrm{S,\uparrow }}\left( \protect%
\varepsilon \right) -f_{\mathrm{D,\uparrow }}\left( \protect\varepsilon %
\right) \right] $ measures the difference at QD level $\protect\varepsilon $%
, where $\Gamma $ is the overall tunnel rate. The difference in $I_{%
\mathrm{D}}$ between \textbf{b} and \textbf{c} with the same $\protect%
\varepsilon -\protect\mu _{\mathrm{D,\uparrow }}$ but different $V_{\mathrm{%
eff}}$ [($=\left( \protect\mu _{\mathrm{D,\uparrow }}-\protect\mu _{%
\mathrm{S,\uparrow }}\right) /e$)] is used to estimate the derivative $%
\frac{d}{dE}f_{\mathrm{S,\uparrow }}$. The integrated current $K_{I}=%
\protect\int I_{\mathrm{D}}d\protect\varepsilon $ measures the area enclosed
by $f_{\mathrm{S,\uparrow }}$ and $f_{\mathrm{D,\uparrow }}$. \textbf{d}, $%
K_{I}$ as a function of $\protect\mu _{\mathrm{D,\uparrow }}$ ($=-eV_{%
\mathrm{D}}$), from which $\protect\mu _{\mathrm{S,\uparrow }}$ is obtained. 
\textbf{e}, The integrated heat currents $K_{J,\protect\varepsilon -\bar{%
\protect\mu}}=\protect\int \left( \protect\varepsilon -\bar{\protect\mu}%
\right) I_{\mathrm{D}}d\protect\varepsilon $ (the red dashed line) and $%
K_{J,\protect\varepsilon _{\mathrm{G}}}=\protect\int \protect\varepsilon _{%
\mathrm{G}}I_{\mathrm{D}}d\protect\varepsilon $ (the red solid line) as
a function of $\protect\mu _{\mathrm{D,\uparrow }}$. $K_{J,\protect%
\varepsilon -\bar{\protect\mu}}$ provides the difference of the heat
currents $\Delta J$ ($=J_{\mathrm{S,\uparrow }}-J_{\mathrm{D,\uparrow }}$)
independent of $\protect\mu _{\mathrm{D,\uparrow }}$, and $K_{J,\protect%
\varepsilon _{\mathrm{G}}}$ provides $\Delta J$ only at $\protect\mu _{%
\mathrm{D,\uparrow }}=\protect\mu _{\mathrm{S,\uparrow }}$, from which $%
\protect\varepsilon -\bar{\protect\mu}$ can be determined.}
\end{figure*}

Data was analyzed by considering the non-thermal heat flow, as summarized in
Fig. 3a. The total heat current $J_{\mathrm{T}}\simeq J_{\mathrm{S},\uparrow
}+J_{\mathrm{S},\downarrow }+J_{\mathrm{E},\uparrow }+J_{\mathrm{E}%
,\downarrow }$ is distributed in the four channels, and the fraction $J_{%
\mathrm{S},\uparrow }\simeq \frac{1}{4}J_{\mathrm{T}}$ in channel C$_{%
\mathrm{S},\uparrow }$ constitutes non-thermal distribution function $f_{%
\mathrm{S},\uparrow }$ in the heat source attached to the QD. We are aware
that the drain channel C$_{\mathrm{D},\uparrow }$ is also heated up to the
heat current $J_{\mathrm{D},\uparrow }$ with a heat transfer factor $\kappa $
due to the long-range interaction \cite{ProkudinaPRL2014,WashioPRB2016}.
Therefore, $f_{\mathrm{S},\uparrow }$, $f_{\mathrm{D},\uparrow }$, and $\mu
_{\mathrm{S},\uparrow }$, as well as $\varepsilon $, have to be determined
to analyze the thermoelectricity.

Because of the negligible level broadening ($w<k_{\mathrm{B}}T_{\mathrm{B}}$%
), $I_{\mathrm{D}}\simeq -e\Gamma \left[ f_{\mathrm{S},\uparrow }\left(
\varepsilon \right) -f_{\mathrm{D},\uparrow }\left( \varepsilon \right) %
\right] $ measures the difference of the distribution functions at QD energy 
$\varepsilon $, as shown in Fig. 3b. The role of $f_{\mathrm{D},\uparrow }$
can be removed by differentiating $I_{\mathrm{D}}$'s taken with the same $f_{%
\mathrm{D},\uparrow }\left( \varepsilon \right) $ at the same $\varepsilon
-\mu _{\mathrm{D}}$ condition (compare Figs. 3b and 3c), from which we
estimate $\frac{d}{dE}f_{\mathrm{S},\uparrow }$ and $f_{\mathrm{S},\uparrow
} $. Similarly, $f_{\mathrm{D},\uparrow }$ can be estimated. We used the
following procedure to do this.

The integrated current $K_{I}=\int I_{\mathrm{D}}d\varepsilon $ over the
current peak should measure $K_{I}=e\Gamma \left( \mu _{\mathrm{D},\uparrow
}-\mu _{\mathrm{S},\uparrow }\right) $, as shown in Fig. 3d. Therefore, $\mu
_{\mathrm{S},\uparrow }$ can be determined from the condition for $K_{I}=0$.
We also consider the integrated heat current $K_{J,\varepsilon -\bar{\mu}%
}=\int \left( \varepsilon -\bar{\mu}\right) I_{\mathrm{D}}d\varepsilon $
over the current peak, where $\bar{\mu}=\left( \mu _{\mathrm{D},\uparrow
}+\mu _{\mathrm{S},\uparrow }\right) /2$ is the average chemical potential.
This $K_{J,\varepsilon -\bar{\mu}}$ measures the difference of the heat
current $\Delta J=J_{\mathrm{S},\uparrow }-J_{\mathrm{D},\uparrow }$
independent of $\mu _{\mathrm{D},\uparrow }$, as shown by the red dashed
line in Fig. 3e. However, we do not know $\varepsilon -\bar{\mu}$ yet,
because $\varepsilon $ ($=\varepsilon _{\mathrm{G}}+\tilde{\varepsilon}$)
deviates from the gate tunable part $\varepsilon _{\mathrm{G}}=\alpha V_{%
\mathrm{GD}}$ with the lever-arm factor $\alpha $ by unknown shift $\tilde{%
\varepsilon}$, which comes from electrostatic shift due to the change of $%
\mu _{\mathrm{D},\uparrow }$ and background charge fractionations.\ Instead,
we calculate $K_{J,\varepsilon _{\mathrm{G}}}=\int \varepsilon _{\mathrm{G}%
}I_{\mathrm{D}}d\varepsilon $, which depends on $\mu _{\mathrm{D},\uparrow }$
and provides $\Delta J$ at $\mu _{\mathrm{D},\uparrow }=\mu _{\mathrm{S}%
,\uparrow }$ (the red solid line). This $K_{J,\varepsilon _{\mathrm{G}}}$ is
used to estimate $\varepsilon -\bar{\mu}$ for each data point at $\left( V_{%
\mathrm{GD}},V_{\mathrm{D}}\right) $ (see Methods).

\begin{figure}[tb]
\begin{center}
\includegraphics[width = 3.3in]{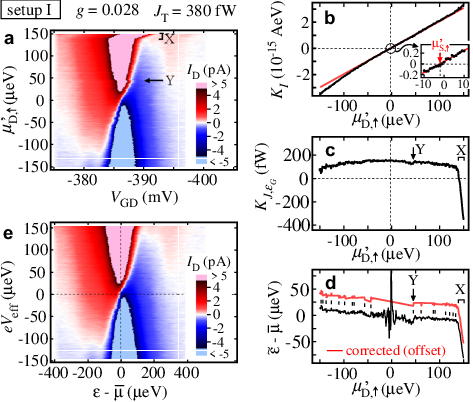}
\end{center}
\caption{\textbf{Conversion from voltage-dependent current }$I_{\mathrm{D%
}}\left( V_{\mathrm{GD}},V_{\mathrm{D}}\right) $\textbf{\ to
energy-dependent current }$I_{\mathrm{D}}\left( \protect\varepsilon -%
\bar{\protect\mu},eV_{\mathrm{eff}}\right) $\textbf{.} \textbf{a}, The
original data $I_{\mathrm{D}}\left( V_{\mathrm{GD}},\protect\mu _{\mathrm{%
D,\uparrow }}^{\prime }\right) $ taken as a function of the gate voltage $%
V_{\mathrm{GD}}$ and the nominal drain potential $\protect\mu _{\mathrm{%
D,\uparrow }}^{\prime }$ providing the actual drain potential $eV_{\mathrm{D}%
}=\protect\mu _{\mathrm{D,\uparrow }}^{\prime }+\protect\mu _{\mathrm{D,0}}$
with\ predicted offset $\protect\mu _{\mathrm{D,0}}\simeq $ 30 $\protect\mu $%
eV in setup I. \textbf{b}, The integrated current $K_{I}$ as a function of 
$\protect\mu _{\mathrm{D,\uparrow }}^{\prime }$. The overall tunnel rate $%
\Gamma $ and the nominal source potential $\protect\mu _{\mathrm{S,\uparrow }%
}^{\prime }$ is obtained from the fit (the red line) with the relation $%
K_{I}=e\Gamma \left( \protect\mu _{\mathrm{D,\uparrow }}^{\prime }-\protect%
\mu _{\mathrm{S,\uparrow }}^{\prime }\right) $. The inset shows the
magnified plot around $K_{I}=0$ (the arrow). \textbf{c}, The energy current $%
K_{J,\protect\varepsilon _{\mathrm{G}}}$ as a function of $\protect\mu _{%
\mathrm{D,\uparrow }}^{\prime }$. The heat-current difference $\Delta J$
is obtained from $K_{J,\protect\varepsilon _{\mathrm{G}}}$ value at $\protect%
\mu _{\mathrm{D,\uparrow }}^{\prime }=\protect\mu _{\mathrm{S,\uparrow }%
}^{\prime }$ (the vertical dashed line). \textbf{d}, The fluctuating
quantum-dot level $\tilde{\protect\varepsilon}$ measured from $\bar{\protect%
\mu}=\left( \protect\mu _{\mathrm{S,\uparrow }}+\protect\mu _{\mathrm{%
D,\uparrow }}\right) /2$, $\tilde{\protect\varepsilon}-\bar{\protect\mu}$,
as a function of $\protect\mu _{\mathrm{D,\uparrow }}^{\prime }$. The
black trace is obtained by using the relation $\tilde{\protect\varepsilon%
}-\bar{\protect\mu}=eh\Gamma \left( K_{J,\protect\varepsilon _{\mathrm{G}%
}}-\Delta J\right) /K_{I}$ with the data $K_{I}$ and $\Gamma $ in 
\textbf{b} and $K_{J,\protect\varepsilon _{\mathrm{G}}}$ and $\Delta J$
in \textbf{c}. The conversion from $\protect\varepsilon _{\mathrm{G}}$ ($%
=-\protect\alpha V_{\mathrm{GD}}$ with the lever-arm factor $\protect\alpha $%
) to $\protect\varepsilon -\bar{\protect\mu}$ ($=\protect\varepsilon _{%
\mathrm{G}}+\tilde{\protect\varepsilon}-\bar{\protect\mu}$) is performed by
the red trace of $\tilde{\protect\varepsilon}-\bar{\protect\mu}$, which was
obtained from the correction of switching events at the vertical bars and
linear fitting between them. The red trace is shifted vertically for
clarity. \textbf{e}, The converted plot of $I_{\mathrm{D}}\left( \protect%
\varepsilon -\bar{\protect\mu},eV_{\mathrm{eff}}\right) $. The initial drift
X and jump Y in \textbf{a} are removed in \textbf{e}.}
\end{figure}

This allows us to convert an original data $I_{\mathrm{D}}\left( V_{\mathrm{%
GD}},V_{\mathrm{D}}\right) $, i.e., $I_{\mathrm{D}}$ taken as a function of $%
V_{\mathrm{GD}}$ and $V_{\mathrm{D}}$, to an energy-dependent data $I_{%
\mathrm{D}}\left( \varepsilon -\bar{\mu},V_{\mathrm{eff}}\right) $. As an
example, $I_{\mathrm{D}}\left( V_{\mathrm{GD}},V_{\mathrm{D}}\right) $ in
Fig. 4a is converted to $I_{\mathrm{D}}\left( \varepsilon -\bar{\mu},V_{%
\mathrm{eff}}\right) $ in Fig. 4e by evaluating $K_{I}$ in Fig. 4b, $%
K_{J,\varepsilon _{\mathrm{G}}}$ in Fig. 4c, and $\tilde{\varepsilon}-\bar{%
\mu}$ in Fig. 4d (see Methods). The generated power $P=-I_{\mathrm{D}}V_{%
\mathrm{eff}}$ is plotted as a function of $\varepsilon -\bar{\mu}$ and $eV_{%
\mathrm{eff}}$ in the color scale of Fig. 2c for the NT state and Fig. 2d
for the QT state. With this conversion, the Coulomb diamond (dashed lines $%
\varepsilon =\mu _{\mathrm{D},\uparrow }$ and $\varepsilon =\mu _{\mathrm{S}%
,\uparrow }$) is symmetrized.

\begin{figure}[tb]
\begin{center}
\includegraphics[width = 3.3in]{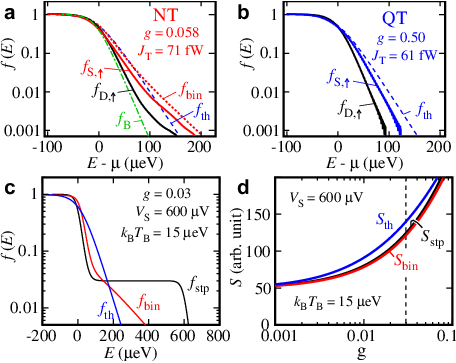}
\end{center}
\caption{\textbf{Energy distribution functions of non-thermal (NT) and
quasi-thermal (QT) states.} \textbf{a}, \textbf{b}, Distribution
functions $f_{\mathrm{S,\uparrow }}(E)$ of the source spin-up channel C$%
_{\mathrm{S},\uparrow }$ and $f_{\mathrm{D,\uparrow }}(E)$ of the drain
spin-up channel C$_{\mathrm{D},\uparrow }$ extracted from the data
obtained with a NT state (the transmission coefficient $g=$ 0.058, the
total heat power $J_{\mathrm{T}}=71$ fW, and the source voltage $V_{%
\mathrm{S}}=$ 183 $\protect\mu $V) in \textbf{a} and a QT state ($g=$ 0.5, $%
J_{\mathrm{T}}=61$ fW, and $V_{\mathrm{S}}=$ 79 $\protect\mu $V) in \textbf{b%
}. The empirical binary Fermi distribution function $f_{\mathrm{bin}}(E)$
with the fraction $p=$ 0.21, the high thermal energy $k_{\mathrm{B}%
}T_{\mathrm{S}}=$ 36 $\protect\mu $eV, and the low thermal energy $k_{%
\mathrm{B}}T_{\mathrm{L}}=$ 15.5 $\protect\mu $eV in \textbf{a}, the
thermalized distribution function $f_{\mathrm{th}}(E)$ with thermal energy
$k_{\mathrm{B}}T_{\mathrm{th}}=$ 22 $\protect\mu $eV in \textbf{a} and 22 $%
\protect\mu $eV in \textbf{b}, and the Fermi distribution function $f_{%
\mathrm{B}}$ at thermal energy $k_{\mathrm{B}}T_{\mathrm{B}}=$ 15 $\protect%
\mu $eV for the base temperature $T_{\mathrm{B}}=$ 175 mK in \textbf{a}
are also shown. \textbf{c}, Comparison of initial double-step distribution
functions $f_{\mathrm{stp}}(E)$, the binary Fermi function $f_{\mathrm{bin}%
}(E)$, and the thermalized function $f_{\mathrm{th}}(E)$ for $g=$ 0.03, $J_{%
\mathrm{T}}=$ 410 fW, and $V_{\mathrm{S}}=$ 600 $\protect\mu $V. \textbf{d},
Comparison of initial entropy $S_{\mathrm{stp}}$ for $f_{\mathrm{stp}}$, the
non-thermal entropy $S_{\mathrm{bin}}$ for $f_{\mathrm{bin}}$, and the
thermalized entropy $S_{\mathrm{th}}$ for $f_{\mathrm{th}}$ as a function of
normalized conductance $g$ of the QPC. $S_{\mathrm{bin}}\simeq S_{\mathrm{%
stp}}$ suggests entropy-conserving equilibration. The dashed line
indicates the condition $g=0.03$ for \textbf{c}. }
\end{figure}

By using the converted data $I_{\mathrm{D}}\left( \varepsilon ,eV_{\mathrm{%
eff}}\right) $, we obtain the derivative $\left. \frac{\partial }{\partial E}%
f_{\mathrm{S}}\right\vert _{E=\varepsilon }\simeq \frac{-1}{e\Gamma \epsilon 
}\left[ I_{\mathrm{D}}\left( \varepsilon +\frac{\epsilon }{4},eV_{\mathrm{eff%
}}+\frac{\epsilon }{2}\right) -I_{\mathrm{D}}\left( \varepsilon -\frac{%
\epsilon }{4},eV_{\mathrm{eff}}-\frac{\epsilon }{2}\right) \right] $ for
small $\epsilon $ ($\lesssim k_{\mathrm{B}}T_{\mathrm{B}}$), and $f_{\mathrm{%
S}}(E)$ is calculated by integrating $\frac{\partial }{\partial E}f_{\mathrm{%
S}}$. $f_{\mathrm{D}}(E)$ can be obtained similarly. With this scheme, $f_{%
\mathrm{S,\uparrow }}(E)$ and $f_{\mathrm{D,\uparrow }}(E)$ in Figs. 5a for
the NT state and Fig. 5b for the QT state are obtained with $\epsilon \simeq 
$ 18 $\mu $eV (comparable to $k_{\mathrm{B}}T_{\mathrm{B}}\simeq $ 15 $\mu $%
eV).

\textbf{Thermoelectricity.} We evaluate fundamental thermoelectric
characteristics from Figs. 2c and 2d. First, electromotive force $V_{\mathrm{%
emf}}$ can be defined as maximum $V_{\mathrm{eff}}$ to produce finite $P>0$.
Data shows higher electromotive force $V_{\mathrm{emf}}$ $\simeq $ 130 $\mu $%
V for the NT state than $V_{\mathrm{emf}}$ $\simeq $ 50 $\mu $V for the QT
state.

Second, the idealized efficiency $\bar{\eta}=eV_{\mathrm{eff}}/\left(
\varepsilon -\mu _{\mathrm{S},\uparrow }\right) $ can be read from the plots 
$I_{\mathrm{D}}\left( \varepsilon -\bar{\mu},eV_{\mathrm{eff}}\right) $. We
draw in Figs. 2c and 2d the solid lines with slopes indicating $\bar{\eta}=$
0.2, 0.4, and 0.6. As these lines indicate, power generation under
conditions closer to the $\varepsilon =\mu _{\mathrm{D},\uparrow }$ line
implies a higher $\bar{\eta}$. Although we do not measure efficiency
directly, the $\bar{\eta}$ values in the $P>0$ region represent the
idealized efficiency that would be achieved if the QD were replaced with an
idealized one. This $\bar{\eta}$ can be used to evaluate the non-thermal
heat source. Defining $\bar{\eta}_{\mathrm{Z}}$ as the maximum $\bar{\eta}$
in the zero power limit ($P=+0$), $\bar{\eta}_{\mathrm{Z}}$ is well beyond
0.6 at large $\left\vert V_{\mathrm{eff}}\right\vert \simeq $ 100 $\mu $V
and $\left\vert \varepsilon \right\vert \simeq $ 100 $\mu $V for the NT
state, while $\bar{\eta}_{\mathrm{Z}}$ is at most 0.6 for the QT state (see
Methods and Supplementary Fig. 4c for systematic error in $\bar{\eta}_{%
\mathrm{Z}}$). This confirms the superior efficiency with the NT state.

The white circles in Figs. 2c and 2d indicate the conditions that yield the
maximum power $P_{\mathrm{M}}$. While $P_{\mathrm{M}}$ is comparable for the
NT and QT states, the corresponding $\bar{\eta}_{\mathrm{M}}$ is greater
than 0.4 for the NT state, but $\bar{\eta}_{\mathrm{M}}<0.4$ for the QT
state. The greater $\bar{\eta}_{\mathrm{M}}$ at the maximum power is
attractive for energy harvesting applications.

\begin{figure*}[tb]
\begin{center}
\includegraphics[width = 6in]{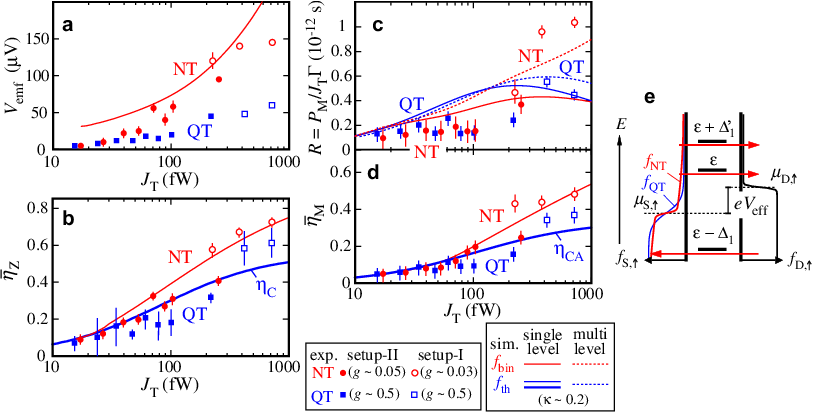}
\end{center}
\caption{ \textbf{Energy-harvesting performances for non-thermal (NT)
and quasi-thermal (QT) states.} \textbf{a}, Electromotive force $V_{%
\mathrm{emf}}$. \textbf{b}, Maximum idealized efficiency $\bar{\protect\eta}%
_{\mathrm{Z}}$ in the zero power limit. \textbf{c}, Maximum normalized
energy-recovery rate $R=P_{\mathrm{M}}/J_{\mathrm{T}}\Gamma $, where
electric power $P_{\mathrm{M}}$ is generated from total waste heat $J_{%
\mathrm{T}}$ by a quantum-dot (QD) heat engine with overall tunnel rate $%
\Gamma $. \textbf{d}, Idealized efficiency $\bar{\protect\eta}_{\mathrm{M}%
} $ at the condition of $R$. Data were taken at transmission coefficient %
$g=0.03-0.05$ (red circles) for NT states and $g=0.3-0.5$ (blue squares) for
QT states obtained with setup I (open symbols) and II (filled symbols).
Error bars represent typical uncertainty in the determination of drain
current $I_{\mathrm{D}}$, effective bias $V_{\mathrm{eff}}$, and QD
level $\protect\varepsilon $ for each data (see Methods). Simulations with
the binary Fermi distribution function $f_{\mathrm{bin}}$ for the NT
state (red lines) and the thermalized Fermi distribution function $f_{%
\mathrm{th}}$ for the QT state (blue lines) are obtained for single- (solid
lines) and multiple-level (dashed lines) QD models. \textbf{e}, Schematic
energy diagram of the QD heat engine in the presence of excited states
with level spacings $\Delta _{1}$ and $\Delta _{1}^{\prime }$. Transport
through the level $\protect\varepsilon +\Delta _{1}^{\prime }$ increases the
thermoelectric current, but that through $\protect\varepsilon -\Delta _{1}$
decreases the current. }
\end{figure*}

We repeated similar experiments at various $J_{\mathrm{T}}$, and the results
from setups I and II are shown by open and filled symbols, respectively, in
Fig. 6. Overall, as compared to the QT states (blue symbols), the NT state
(red symbols) shows larger $V_{\mathrm{emf}}$ in Fig. 6a, larger $\bar{\eta}%
_{\mathrm{Z}}$ in Fig. 6b, and larger $\bar{\eta}_{\mathrm{M}}$ in Fig. 6d.
The advantage of non-thermal states are confirmed in both setups.

We also evaluate the heat recovery efficiency $P_{\mathrm{M}}/J_{\mathrm{T}}$%
, which measures how much fraction of waste heat $J_{\mathrm{T}}$ can be
reconverted to electric power $P_{\mathrm{M}}$. Since $P_{\mathrm{M}}$
depends linearly on $\Gamma $ in the small $w$ limit, the normalized value $%
R=P_{\mathrm{M}}/J_{\mathrm{T}}\Gamma $ is plotted in Fig. 6c to compare the
data taken with different $\Gamma $. Because the heat of the NT state is
distributed over a wide energy range, NT states should provide smaller $R$,
as seen in the experimental data of setup II (filled symbols). In setup I
(open symbols), however, NT states show larger $R$ possibly with the help of
the excited state, as we discuss in the simulation.

\begin{figure*}[tb]
\begin{center}
\includegraphics[width = 7in]{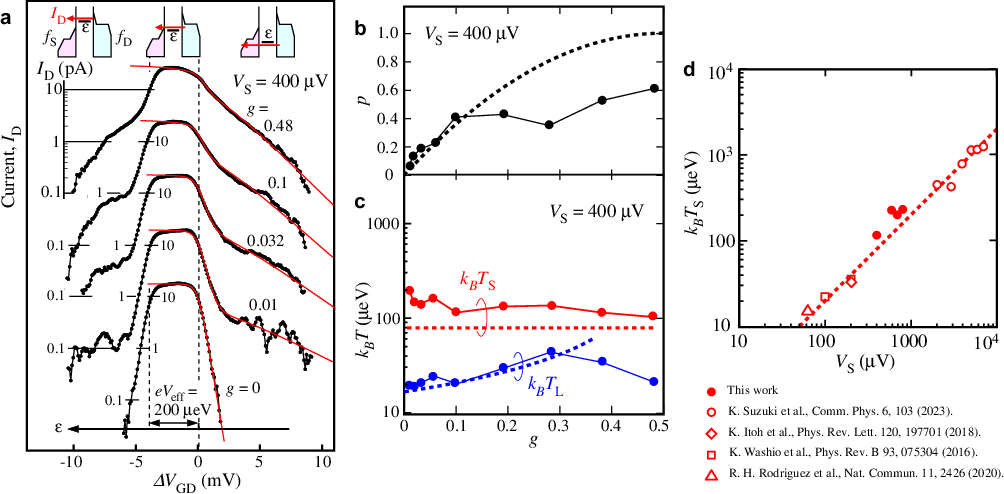}
\end{center}
\caption{\textbf{Binary Fermi distribution function }$f_{\mathrm{bin}}$. 
\textbf{a}, Current ($I_{\mathrm{D}}$) spectra taken at various
transmission coefficient $g$ and a source bias voltage $V_{\mathrm{S}%
}=$ 400 $\protect\mu $V in setup I. The current measures the energy
distribution function $1-f_{\mathrm{S},\uparrow }$ of \ the source channel
C$_{\mathrm{S},\uparrow }$ on the right side and $f_{\mathrm{D},\uparrow }$
of the drain channel C$_{\mathrm{D},\uparrow }$ on the left side of the
peak. The insets show corresponding energy diagrams. The red lines show $%
1-f_{\mathrm{bin}}$ fitted to the data. \textbf{b} and \textbf{c}, $g$
dependence of the fitting parameters (the fraction $p$, the high thermal
energy $k_{\mathrm{B}}T_{\mathrm{S}}$, the low thermal energy $k_{\mathrm{B}%
}T_{\mathrm{L}}$) for the data at $V_{\mathrm{S}}=$ 400 $\protect\mu $V. 
\textbf{d}, $V_{\mathrm{S}}$ dependence of $k_{\mathrm{B}}T_{\mathrm{S}}$
obtained in the present device (the solid circles) and analyzed from the
data in the literature ($\bigtriangleup $ for Ref. \protect\cite%
{RodriguezNatComm2020}, $\square $ for Ref. \protect\cite{WashioPRB2016}, $%
\Diamond $ for Ref. \protect\cite{Itoh-PRL2018}, and $\bigcirc $ for Ref. 
\protect\cite{SuzukiCommPhys2023}). The dashed lines in \textbf{b }- \textbf{%
d} show the proposed parameters, $p=4g(1-g)$, $k_{\mathrm{B}}T_{\mathrm{S}}=%
\protect\sqrt{\frac{3}{8\protect\pi ^{2}}}eV_{\mathrm{S}}$, and $k_{\mathrm{B%
}}T_{\mathrm{L}}=k_{\mathrm{B}}T_{\mathrm{B}}/\protect\sqrt{1-p}$ for
thermal energy $k_{\mathrm{B}}T_{\mathrm{B}}=$ 15 $\protect\mu $eV at base
temperature $T_{\mathrm{B}}$. }
\end{figure*}

\textbf{Non-thermal state.} The above characteristics should be attributed
to the non-thermal distribution function $f_{\mathrm{NT}}(E)$. Because
theoretical derivation of $f_{\mathrm{NT}}(E)$ is cumbersome \cite%
{LevkivskyiPRB2012,vonDelftAnnPhys1998,FujisawaAnnPhys2022}, we adopt
empirical binary Fermi distribution function $f_{\mathrm{bin}}\left( E;p,T_{%
\mathrm{S}},T_{\mathrm{L}}\right) =pf_{\mathrm{FD}}\left( E;\mu _{\mathrm{S}%
,\uparrow },T_{\mathrm{S}}\right) +\left( 1-p\right) f_{\mathrm{FD}}\left(
E;\mu _{\mathrm{S},\uparrow },T_{\mathrm{L}}\right) $. Here, $f_{\mathrm{FD}%
}(E;\mu ,T)$ is the Fermi-Dirac distribution function with chemical
potential $\mu $ and temperature $T$. This $f_{\mathrm{bin}}$ reproduces
most of the experiments in the literature \cite%
{WashioPRB2016,Itoh-PRL2018,RodriguezNatComm2020,SuzukiCommPhys2023}. We
reconfirmed this $f_{\mathrm{bin}}$ by using the same QD at large $V_{%
\mathrm{eff}}=$ 200 $\mu $V, as shown in Fig. 7a. The bottom trace obtained
without heat injection ($g=0$) shows normal exponential tails on both sides
of the current peak, from which the base electron temperature $T_{\mathrm{B}%
}\simeq $ 150 mK is estimated. Under heat injection ($g>0$), the peak
exhibits non-exponential tails on both sides. The right tail measures the
hole excitation of the source, as illustrated in the right inset. The data
can be fitted nicely by $f_{\mathrm{bin}}\left( E;p,T_{\mathrm{S}},T_{%
\mathrm{L}}\right) $, as shown by the red line. The fitted parameters ($p$, $%
k_{\mathrm{B}}T_{\mathrm{S}}$, and $k_{\mathrm{B}}T_{\mathrm{L}}$) are
plotted as a function of $g$ in Figs. 7b and 7c. One can see that $p$
increases with $g$, $k_{\mathrm{B}}T_{\mathrm{S}}$ is independent of $g$,
and $k_{\mathrm{B}}T_{\mathrm{L}}$ gradually increases with $g$. Figure 7d
shows that $k_{\mathrm{B}}T_{\mathrm{S}}$ increases linearly with $V_{%
\mathrm{S}}$, where the values extracted from the literature are also
plotted.

These experimental data suggest us to use $p=4g(1-g)$, $k_{\mathrm{B}}T_{%
\mathrm{S}}=\sqrt{\frac{3}{8\pi ^{2}}}eV_{\mathrm{S}}$, and $T_{\mathrm{L}%
}=T_{\mathrm{B}}/\sqrt{1-p}$ for small $g\ll \frac{1}{2}$, as shown by the
dashed lines in Figs. 7b-d. Namely, $p$ depends only on $g$, and $T_{\mathrm{%
S}}$ depends only on $V_{\mathrm{S}}$. $T_{\mathrm{L}}$ is adjusted to yield
heat current $J_{\mathrm{S},\uparrow }=\frac{1}{4}J_{\mathrm{T}}+J_{\mathrm{B%
}}$ by assuming the equal heat distribution among the four channels (C$_{%
\mathrm{E},\uparrow }$, C$_{\mathrm{E},\downarrow }$, C$_{\mathrm{S}%
,\uparrow }$, and C$_{\mathrm{S},\downarrow }$) on the original heat current 
$J_{\mathrm{B}}=\frac{\pi ^{2}}{6h}\left( k_{\mathrm{B}}T_{\mathrm{B}%
}\right) ^{2}$ at $T_{\mathrm{B}}$. $f_{\mathrm{bin}}$ is determined from
the heat injection ($g$ and $V_{\mathrm{S}}$) without any free parameters
and is used in the following simulations. The binary Fermi distribution
function becomes unclear at $g\simeq $ 0.5, as seen in the topmost trace of
Fig. 7a, where the channel can be regarded as in the QT state.

We have extracted the distribution functions $f_{\mathrm{S},\uparrow }(E)$
and $f_{\mathrm{D},\uparrow }(E)$ in Fig. 5a for the NT state and Fig. 5b
for the QT state. As compared to the QT state with a single exponential
decay in $f_{\mathrm{S,\uparrow }}(E)$ at $E-\mu \gtrsim $ 0 of Fig. 5b, the
NT state in Fig. 5a shows an additional gentle decay in $f_{\mathrm{%
S,\uparrow }}(E)$ at $E-\mu \gtrsim $ 100 $\mu $eV. This $f_{\mathrm{%
S,\uparrow }}(E)$ is close to the above $f_{\mathrm{bin}}$ shown by the
dotted line. If the electrons were fully thermalized, the Fermi distribution
function $f_{\mathrm{th}}\left( E\right) =f_{\mathrm{FD}}(E;\mu _{\mathrm{S}%
,\uparrow },T_{\mathrm{th}})$ would appear with the thermalized temperature $%
T_{\mathrm{th}}=\sqrt{\frac{3h}{2\pi ^{2}k_{\mathrm{B}}^{2}}J_{\mathrm{T}%
}+T_{\mathrm{B}}^{2}}$. This $f_{\mathrm{th}}\left( E\right) $ is closer to
the extracted $f_{\mathrm{S,\uparrow }}(E)$ for the QT state in Fig. 5b,
indicating that the QT state can be regarded as a thermalized reference
state.

The drain channel is also heated, though only slightly, as indicated by the
smaller magnitude of the gently decaying component of $f_{\mathrm{D}%
,\uparrow }(E)$ in Fig. 5a, possibly due to the long-range interaction
between C$_{\mathrm{S},\uparrow }$ and C$_{\mathrm{D},\uparrow }$ \cite%
{ProkudinaPRL2014,WashioPRB2016,HashisakaPRB2012}. In the following
simulations, we assume that a fraction $\kappa \simeq 0.2$ of heat in C$_{%
\mathrm{S},\uparrow }$ is transferred to C$_{\mathrm{D},\uparrow }$ to
induce non-thermal distribution function $f_{\mathrm{D},\uparrow }(E)=f_{%
\mathrm{bin}}\left( E;\kappa p,T_{\mathrm{S}},T_{\mathrm{L}}\right) $ for
the NT state and thermal one $f_{\mathrm{D},\uparrow }(E)=f_{\mathrm{FD}%
}(E;\mu _{\mathrm{S},\uparrow },T_{\mathrm{th}}^{\prime })$ with $T_{\mathrm{%
th}}^{\prime }=\sqrt{\frac{3h}{2\pi ^{2}k_{\mathrm{B}}^{2}}\kappa J_{\mathrm{%
T}}+T_{\mathrm{B}}^{2}}$ for the QT state.

Importantly, the binary Fermi distribution function is consistent with the
entropy argument. Because spin and charge quasi-particles in TL liquids are
non-interacting, the entropy should not change from the initial state. When
electrons are partitioned at the QPC with energy-independent $g$, the
initial distribution function in C$_{\mathrm{S},\uparrow }$ should be a
double-step function $f_{\mathrm{stp}}(E)=(1-g)f_{\mathrm{FD}}(E;\mu _{%
\mathrm{S},\uparrow }-geV_{\mathrm{S}},T_{\mathrm{B}})+gf_{\mathrm{FD}%
}(E;\mu _{\mathrm{S},\uparrow }+(1-g)eV_{\mathrm{S}},T_{\mathrm{B}})$ with
the base temperature $T_{\mathrm{B}}$. This $f_{\mathrm{stp}}(E)$ is
short-lived due to the electron-electron interaction, and the system results
in a long-lived non-thermal state \cite{WashioPRB2016,Itoh-PRL2018}. The
system remains in the NT state with $f_{\mathrm{bin}}$ for a long distance,
rather than relaxing into the thermalized state with $f_{\mathrm{th}}$.
Distribution functions $f_{\mathrm{stp}}(E)$, $f_{\mathrm{bin}}\left(
E\right) $, and $f_{\mathrm{th}}\left( E\right) $ for $g=0.03$ and $V_{%
\mathrm{S}}=$ 600 $\mu $V are shown in Fig. 5c. Corresponding entropies $S_{%
\mathrm{stp}}$ for $f_{\mathrm{stp}}$, $S_{\mathrm{bin}}$ for $f_{\mathrm{bin%
}}$, and $S_{\mathrm{th}}$ for $f_{\mathrm{th}}$ are plotted as a function
of $g$ in Fig. 5d, where the entropy $S=-\frac{k_{\mathrm{B}}}{hv}\int [f\ln
f+(1-f)\ln (1-f)]dE$ is obtained from the distribution function $f$ for a
constant velocity $v$ of electrons in the channels. The non-thermalizing
nature of the TL liquid is evident from $S_{\mathrm{stp}}$ and $S_{\mathrm{%
bin}}$ being comparable to each other and smaller than $S_{\mathrm{th}}$
particularly in the range of $0.005<g<0.05$, where the experiments were
conducted. The empirical function $f_{\mathrm{bin}}$ captures the isentropic
process of spin-charge separation.

\textbf{Simulations.}

The heat-engine characteristics are qualitatively reproduced in the
simulation based on the idealized single-level QD. This is justified for
setup II ($w<k_{\mathrm{B}}T_{\mathrm{B}}$ and $k_{\mathrm{B}}T_{\mathrm{S}%
}<\Delta _{1},\Delta _{1}^{\prime }$) even at largest $J_{\mathrm{T}}\simeq $
250 fW ($k_{\mathrm{B}}T_{\mathrm{S}}\simeq $ 70 $\mu $eV smaller than $%
\Delta _{1}\simeq \Delta _{1}^{\prime }\simeq $ 400 $\mu $eV). Excited
states in setup I play minor roles in the estimate of $V_{\mathrm{emf}}$ and 
$\bar{\eta}$'s, as discussed below. For distribution functions, we use $f_{%
\mathrm{S},\uparrow }(E)=f_{\mathrm{bin}}\left( E;p,T_{\mathrm{S}},T_{%
\mathrm{L}}\right) $ and $f_{\mathrm{D},\uparrow }(E)=f_{\mathrm{bin}}\left(
E;\kappa p,T_{\mathrm{S}},T_{\mathrm{L}}\right) $ for NT states and $f_{%
\mathrm{S},\uparrow }(E)=f_{\mathrm{FD}}(E;\mu ,T_{\mathrm{th}})$ and $f_{%
\mathrm{D},\uparrow }(E)=f_{\mathrm{FD}}(E;\mu ,T_{\mathrm{th}}^{\prime })$
for QT states. All parameters ($p$, $T_{\mathrm{S}}$, $T_{\mathrm{L}}$, $T_{%
\mathrm{th}}$, and $T_{\mathrm{th}}^{\prime }$) are determined from the
above formulas with $k_{\mathrm{B}}T_{\mathrm{B}}=$ 15 $\mu $eV and $\kappa
=0.2$ for each heating condition ($g$ and $J_{\mathrm{T}}$). The current $I_{%
\mathrm{D}}=e\Gamma \left[ f_{\mathrm{D},\uparrow }\left( \varepsilon
\right) -f_{\mathrm{S},\uparrow }\left( \varepsilon \right) \right] $ and
the power $P=-I_{\mathrm{D}}V_{\mathrm{eff}}$ are calculated as a function
of $\varepsilon -\bar{\mu}$ and $V_{\mathrm{eff}}$, as shown in Figs. 2e for
the NT state and 2f for the thermalized state under equal $J_{\mathrm{T}}=$
400 fW.

For the thermalized state in Fig. 2f, the maximum $\bar{\eta}$ in the zero
power limit is nothing but the Carnot efficiency $\eta _{\mathrm{C}}=1-T_{%
\mathrm{th}}^{\prime }/T_{\mathrm{th}}$, and the $\bar{\eta}$ at the maximum
power (the white circle) is the Curzon--Ahlborn efficiency $\eta _{\mathrm{CA%
}}=1-\sqrt{T_{\mathrm{th}}^{\prime }/T_{\mathrm{th}}}$ in the thermodynamic
limit \cite{EspositoEuroPhysLett2009}. The non-thermal state in Fig. 2e
shows larger $\bar{\eta}_{\mathrm{Z}}$ ($>\eta _{\mathrm{C}}$) and larger $%
\bar{\eta}_{\mathrm{M}}$ ($>\eta _{\mathrm{CA}}$). The simulation reproduces
the experimental features (larger $\bar{\eta}_{\mathrm{Z}}$ and $\bar{\eta}_{%
\mathrm{M}}$ in Fig. 2c than those in Fig. 2d). In addition, the calculated $%
V_{\mathrm{emf}}\simeq $ 150 $\mu $V in Fig. 2e is similar to the
experimental $V_{\mathrm{emf}}\simeq $ 130 $\mu $V in Fig. 2c. Whereas the $%
V_{\mathrm{emf}}$ calculated for thermalized states in Fig. 2f is infinite,
the experimental $V_{\mathrm{emf}}\simeq $ 50 $\mu $V in Fig. 2d is
practically determined by some unwanted currents. Nevertheless, the
idealized single-level model qualitatively reproduces $V_{\mathrm{emf}}$, $%
\bar{\eta}_{\mathrm{Z}}$, $\bar{\eta}_{\mathrm{M}}$, and $R$, as also shown
by the solid lines in Figs. 6a-6d.

Excited states in the QD induce non-ideal current that can degrade or
enhance the thermoelectric characteristics. For example, consider the energy
diagram ($\varepsilon >\mu _{\mathrm{D},\uparrow }>\mu _{\mathrm{S},\uparrow
}$) in Fig. 6e, where the transport through the ground state $\varepsilon $
primarily induces thermoelectric effect. Transport through the excited state
(electrochemical potential $\varepsilon +\Delta _{1}^{\prime }$) of $\left(
N+1\right) $-electron QD increases $P$, and that ($\varepsilon -\Delta _{1}$%
) of $N$-electron QD decreases $P$. Because $\Delta _{1}^{\prime }\simeq $
80 $\mu $eV is smaller than $\Delta _{1}\simeq $ 400 $\mu $eV in setup I,
the $\left( N+1\right) $-electron excited state might contribute dominantly
to increase $P$. We employed the standard master equation to calculate the
thermoelectricity in the presence of excited states (see Methods). As shown
by the dashed lines in Fig. 6c, the multi-level simulation indicates that $%
R=P_{\mathrm{M}}/J_{\mathrm{T}}\Gamma $ for NT states surpasses $R$ for
thermalized states at larger $J_{\mathrm{T}}$. The simulation also suggests
that $P_{\mathrm{M}}$ can be maximized by tailoring the energy transmission
function of the heat engine \cite{WhitneyPRL2014}. Namely, energy harvesting
with $N=0$ QDs should be advantageous for maximizing $P_{\mathrm{M}}$. While
the excited states modify $P$, the simulation suggests that the
power-generation condition ($P>0$) in the $\left( \varepsilon -\bar{\mu},eV_{%
\mathrm{eff}}\right) $ plane does not change significantly, as it is
primarily determined by the ground-state transport. Therefore, we use the
single-level model to simulate $V_{\mathrm{emf}}$ and $\bar{\eta}$'s even
for setup I.

\bigskip\ 

\begin{figure*}[tb]
\begin{center}
\includegraphics[width = 2.0in]{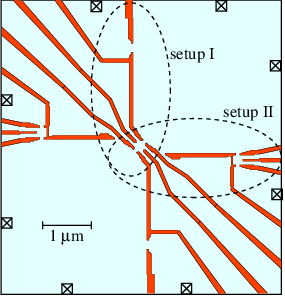}
\end{center}
\caption{\textbf{Gate pattern of the device.} The fine gate pattern for
electron-beam lithography is shown by orange. The dashed lines show the
regions used for setups I and II.}
\end{figure*}

\begin{figure*}[tb]
\begin{center}
\includegraphics[width = 6.4in]{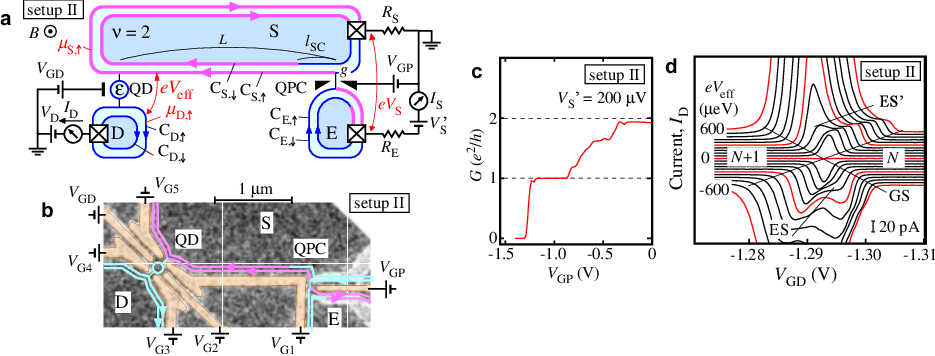}
\end{center}
\caption{\textbf{Setup II.} \textbf{a} Schematic device structure with
emitter (E), source (S), and drain (D) regions. The chirality of the edge
channels and the polarities of the magnetic field $B$ and the supply voltage 
$V_{\mathrm{S}}^{\prime }$ differ from those in setup I. Electrons and heat
travel unidirectionally in the edge channels C$_{r,\uparrow }$ and C$%
_{r,\downarrow }$ of $r\in \left\{ \mathrm{E,S,D}\right\} $. The quantum
point contact (QPC) with the effective bias voltage $V_{\mathrm{S}}$ and the
dimensionless transmission coefficient $g$ generates total heat current $J_{%
\mathrm{T}}$ in the four channels (magenta lines). The quantum dot (QD) with
an effective bias voltage, $V_{\mathrm{eff}}$, works as a heat engine. 
\textbf{b}, Scanning electron micrograph of a control device with false
color for setup II. \textbf{c}, Quantized conductance of the QPC. \textbf{d}%
, Coulomb diamond characteristics of the QD. Each trace is offset for
clarity. Current steps are associated with the ground state (GS) and excited
states (ES and ES').}
\end{figure*}

\begin{figure*}[tb]
\begin{center}
\includegraphics[width = 4.4in]{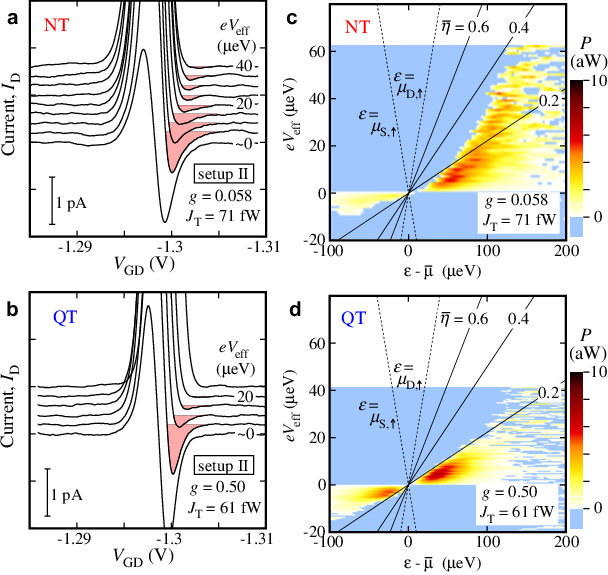}
\end{center}
\caption{\textbf{Heat-engine characteristics with setup II.} \textbf{a}, 
\textbf{b}, Drain current $I_{\mathrm{D}}$ as a function of the gate voltage 
$V_{\mathrm{GD}}$ at various effective energy bias $eV_{\mathrm{eff}}$.
Positive power generation is highlighted by red. \textbf{c}, \textbf{d},
Color plot of the electric power $P$ ($=-I_{\mathrm{D}}V_{\mathrm{eff}}$) as
a function of $eV_{\mathrm{eff}}$ and the dot level $\protect\varepsilon $
relative to the average chemical potential $\bar{\protect\mu}=\left( \protect%
\mu _{\mathrm{S},\uparrow }+\protect\mu _{\mathrm{D},\uparrow }\right) /2$
of the source and drain channels, $\protect\varepsilon -\bar{\protect\mu}$.
Comparable heat power $J_{\mathrm{T}}\simeq $ 60 - 70 fW was used in \textbf{%
a} and \textbf{c} for the NT state, and \textbf{b} and \textbf{d} for the QT
state. The conditions for $\protect\varepsilon =\protect\mu _{\mathrm{D}%
,\uparrow }$ and $\protect\varepsilon =\protect\mu _{\mathrm{S},\uparrow }$,
and the idealized efficiency $\bar{\protect\eta}=$ 0.2, 0.4, and 0.6 are
shown by the dashed and solid lines, respectively, in \textbf{c }and \textbf{%
d}.}
\end{figure*}

\begin{figure*}[tb]
\begin{center}
\includegraphics[width = 4.4in]{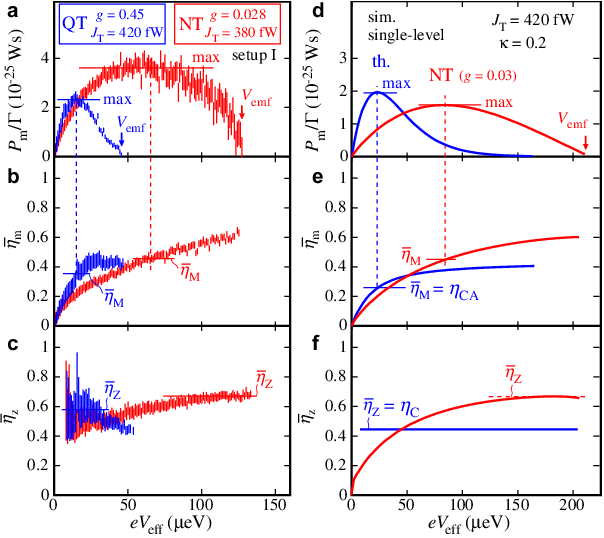}
\end{center}
\caption{\textbf{Estimation of heat-engine performances.} \textbf{a}, The
maximum constrained power $P_{\mathrm{m}}$ normalized by the overall tunnel
rate $\Gamma $, $P_{\mathrm{m}}/\Gamma $, as a function of the effective
energy bias $eV_{\mathrm{eff}}$ obtained from the energy dependent current $%
I_{\mathrm{D}}\left( \protect\varepsilon -\bar{\protect\mu},eV_{\mathrm{eff}%
}\right) $, where $P_{\mathrm{m}}$ is the constrained maximum of $P=-I_{%
\mathrm{D}}V_{\mathrm{eff}}$ when only the dot level $\protect\varepsilon -%
\bar{\protect\mu}$ is changed but $eV_{\mathrm{eff}}$ is fixed. The red and
blue data are obtained from non-thermal (NT) and quasi-thermal (QT) states,
respectively. The maximum value $P_{\mathrm{M}}$ is obtained from the
maximum value of $P_{\mathrm{m}}$ (the horizontal bars labeled `max'). The
electromotive force $V_{\mathrm{emf}}$ is obtained from the vanishing $P_{%
\mathrm{m}}$ (the arrows show $V_{\mathrm{emf}}\simeq $ 127 $\protect\mu $V
for the NT state, and $V_{\mathrm{emf}}\simeq $ 46 $\protect\mu $V for the
QT state). \textbf{b}, The idealized efficiency $\bar{\protect\eta}_{\mathrm{%
m}}$ at the constrained maximum. The efficiency $\bar{\protect\eta}_{\mathrm{%
M}}$ at maximum $P_{\mathrm{M}}$ is obtained from this plot ($\bar{\protect%
\eta}_{\mathrm{M}}\simeq $ 0.45 for the NT state and $\bar{\protect\eta}_{%
\mathrm{M}}\simeq $ 0.35 for the QT state). \textbf{c}, The efficiency $\bar{%
\protect\eta}_{\mathrm{z}}$ in the zero power limit as a function of $eV_{%
\mathrm{eff}}$. The horizontal bars shows $\bar{\protect\eta}_{\mathrm{Z}%
}\simeq $ 0.65 for the NT state and $\bar{\protect\eta}_{\mathrm{M}}\simeq $
0.58 for the QT state. The red and blue data in a - c are obtained from the
NT state ($g=$ 0.028 $J_{\mathrm{T}}=$ 380 fW) and QT state ($g=$ 0.45 $J_{%
\mathrm{T}}=$ 420 fW), respectively, in setup I. \textbf{d}, Simulated $P_{%
\mathrm{m}}/\Gamma $ as a function of $eV_{\mathrm{eff}}$. \textbf{e},
Simulated efficiency $\bar{\protect\eta}_{\mathrm{m}}$ at the constrained
maximum. \textbf{f}, Simulated efficiency $\bar{\protect\eta}_{\mathrm{z}}$
in the zero power limit. The simulations in c-e are obtained by using the
single-level model for the NT state with the binary Fermi distribution
functions at $g=$ 0.03 and $J_{\mathrm{T}}=$ 420 fW (the red lines) and the
corresponding thermalized state with the thermalized Fermi distribution
functions (the blue lines). $\bar{\protect\eta}_{\mathrm{M}}$ and $\bar{%
\protect\eta}_{\mathrm{Z}}$ calculated for the thermalized case are
identical to the Curzon-Ahlborn efficiency $\protect\eta _{\mathrm{CA}}$ and
the Carnot efficiency $\protect\eta _{\mathrm{C}}$, respectively.}
\end{figure*}

{\large Conclusions.}

In summary, non-thermal TL liquid is an attractive working fluid for
heat-energy conversion in energy harvesting. The experiment and simulation
show that non-thermal states yield higher electromotive force $V_{\mathrm{emf%
}}$, higher conversion efficiency $\bar{\eta}_{\mathrm{Z}}$ in the zero
power limit, and higher conversion efficiency $\bar{\eta}_{\mathrm{M}}$ at
maximum power, as compared to the thermalized state. Moreover, non-thermal
states can yield higher heat recovery efficiency $R$ by tailoring the
energy filtering function to collect high-energy electrons \cite%
{WhitneyPRL2014}. The binary Fermi distribution function with the proposed
parameters is useful for estimating the performance of the non-thermal
states. The scheme and the idea can be applied to more general
non-equilibrium states \cite{SanchezPRL2019}, other TL systems \cite%
{TaruchaSSC1995,BookGiamarchi,BockrathNature1999,ChangRMB,AuslaenderScience2005}
and potentially other integrable systems \cite{UedaNRP2020}.

\bigskip

{\large Methods}

\textbf{Device.} The device was fabricated in a standard AlGaAs/GaAs
heterostructure with a two-dimensional electron system (2DES) located at
depth $d=$ 100 nm below the surface. The as-grown 2DES has an electron
density of $n_{\mathrm{e}}=$ 3.1$\times $10$^{11}$ cm$^{-2}$ and
low-temperature mobility of $\mu _{\mathrm{e}}\gtrsim $ 10$^{6}$ cm$^{2}$%
/Vs. The mesoscopic quantum Hall device has several gate electrodes to
selectively activate quantum dots (QDs) and quantum point contacts (QPCs).
The fine gate pattern for electron-beam lithography is shown in
Supplementary Fig. 1. The measurements with setups I and II were performed
in different cooldowns with different QPCs and different characteristics of
the same QD. All measurements were performed at magnetic field $B=$ 6 T ($%
\nu =$ 2) and $T_{\mathrm{B}}\simeq $ 150 mK.

The details of setup I are shown in Fig. 1. Those of setup II are shown in
Supplementary Fig. 2 with the schematic in 2a, scanning electron micrograph
of a control device with false color in 2b, quantized conductance of the QPC
in 2c, and Coulomb diamond characteristics of the QD in 2d. The lithographic
distance between the QD and the QPC is $L=$ 2 $\mu $m for setup I and $L=$
2.1 $\mu $m for setup II.

The QPC conductance $G=I_{\mathrm{S}}/V_{\mathrm{S}}$ is obtained by
measuring current $I_{\mathrm{S}}$ for the effective bias voltage $V_{%
\mathrm{S}}=V_{\mathrm{S}}^{\prime }-\left( R_{\mathrm{S}}+R_{\mathrm{E}%
}\right) I_{\mathrm{S}}$ under the supply voltage $V_{\mathrm{S}}^{\prime }$
and the series resistance $R_{\mathrm{S}}+R_{\mathrm{E}}\simeq $ 33 k$\Omega 
$ in setup I and 12 k$\Omega $ in II. Clear quantized conductance $%
G=ge^{2}/h $ particularly at $g=$ 1 are observed for both setups. The
experiments were performed in the tunneling regime at $g=0.03-0.05$ and $%
\simeq 0.5$ to prepare NT and QT states, respectively, to generate the total
heat power $J_{\mathrm{T}}=\frac{e^{2}}{h}g(1-g)V_{\mathrm{S}}^{2}$. The
heat-engine characteristics for NT and QT states are compared with
comparable $J_{\mathrm{T}}$ by adjusting $V_{\mathrm{S}}$.

The charging energy of the QD is about 1.2 meV for both setups. The
tunneling rates $\Gamma _{\mathrm{S}}$ and $\Gamma _{\mathrm{D}}$ on the
source and drain side, respectively, for the ground state (GS) transport are
made asymmetric with $\Gamma _{\mathrm{S}}\simeq $ 0.16 GHz and $\Gamma _{%
\mathrm{D}}\simeq $ 5.7 GHz for setup I, and symmetric with $\Gamma _{%
\mathrm{S}}\simeq \Gamma _{\mathrm{D}}\simeq $ 1 GHz for setup II. The main
difference between the two QDs appears in the characteristics of excited
states. The QD in setup II shows large $\Delta _{1}$ $\simeq $ $\Delta
_{1}^{\prime }$ $\simeq $ 400 $\mu $eV as compared to $\Delta _{1}^{\prime }$
$\simeq $ 80 $\mu $eV in setup I. However, Supplementary Fig. 2d for setup
II shows vague current steps (the solid line labeled `ES') at $V_{\mathrm{eff%
}}<0$, which suggest large tunneling rate (large lifetime broadening) that
depends strongly on $V_{\mathrm{eff}}$. This large broadening might be the
reason why the heat-energy conversion is not efficient at $V_{\mathrm{eff}%
}<0 $ in setup II (Supplementary Fig. 3). Otherwise, the conversion
characteristics at $V_{\mathrm{eff}}>0$ are similar for both setups, as
summarized in Fig. 6.

\bigskip

\textbf{Representative data for setup II.} We present heat-engine
characteristics obtained with setup II in Supplementary Fig. 3 with line
plots of $I_{\mathrm{D}}$ in 3a for the NT state prepared at $g=$ 0.058 and $%
J_{\mathrm{T}}=$ 71 fW and 3b for the QT state prepared at $g=$ 0.50 and $J_{%
\mathrm{T}}=$ 61 fW. Positive power generation ($I_{\mathrm{D}}<0$ at $V_{%
\mathrm{eff}}>0$) is highlighted by red. Under these comparable total
powers, positive power generation is seen even at large $eV_{\mathrm{eff}}>$
40 $\mu $eV for the NT state but is limited to $eV_{\mathrm{eff}}<$ 25 $\mu $%
eV for the QT state. This comparison confirms that larger electromotive
force can be achieved with the NT state.

We evaluate the heat-engine performances by converting the gate voltage $V_{%
\mathrm{GD}}$ to $\varepsilon -\bar{\mu}$, based on the procedure described
in the main text. The power $P$ ($>0$) is plotted in the\ color scale as a
function of $eV_{\mathrm{eff}}$ and $\varepsilon -\bar{\mu}$ in
Supplementary Fig. 3c for the NT state and Fig. 3d for the QT state. The
conditions for the idealized efficiency $\bar{\eta}=$ 0.2, 0.4, and 0.6 are
shown by the solid lines with the relation $\bar{\eta}=eV_{\mathrm{eff}%
}/\left( \varepsilon -\mu _{\mathrm{S}}\right) $. Positive power generation (%
$P>0$) can be obtained with a large $\bar{\eta}$ beyond 0.2 for the NT state
but with $\bar{\eta}\lesssim 0.2$ for the QT state. In this way, more
efficient heat-energy conversion can be achieved with the NT state.

\bigskip

\textbf{Conversion from }$I_{\mathrm{D}}\left( V_{\mathrm{GD}},V_{\mathrm{D}%
}\right) $\textbf{\ to }$I_{\mathrm{D}}\left( \varepsilon -\bar{\mu},eV_{%
\mathrm{eff}}\right) $\textbf{.} The QD current $I_{\mathrm{D}}\left( V_{%
\mathrm{GD}},V_{\mathrm{D}}\right) $, which was measured by sweeping $V_{%
\mathrm{GD}}$ at various $V_{\mathrm{D}}$ values, is converted to the energy
dependent current $I_{\mathrm{D}}\left( \varepsilon -\bar{\mu},eV_{\mathrm{%
eff}}\right) $ in the following way. While $\mu _{\mathrm{D,\uparrow }}=eV_{%
\mathrm{D}}$ ($e>0$) is given by the drain voltage $-V_{\mathrm{D}}$ shown
in Fig. 1b, we need to precisely determine the chemical potential $\mu _{%
\mathrm{S,\uparrow }}$ of the source and the QD level $\varepsilon $, where $%
\varepsilon $ ($=$ $\varepsilon _{\mathrm{G}}+\tilde{\varepsilon}$) can be
shifted by unknown offset $\tilde{\varepsilon}$ from the gate-tunable part $%
\varepsilon _{\mathrm{G}}=-\alpha V_{\mathrm{GD}}$.

For simplicity, we assume that $\tilde{\varepsilon}$ is fixed during each
sweep of $V_{\mathrm{GD}}$, while $\tilde{\varepsilon}$ may change for
different $V_{\mathrm{D}}$ values. In addition, single-level transport with
negligible level broadening is assumed to provide the relation $I_{\mathrm{D}%
}=e\Gamma \left[ f_{\mathrm{D,\uparrow }}\left( \varepsilon \right) -f_{%
\mathrm{S,\uparrow }}\left( \varepsilon \right) \right] $ with unknown $%
\Gamma $. First, we evaluate integrated current $K_{I}\left( \mu _{\mathrm{%
D,\uparrow }}\right) =\int I_{\mathrm{D}}d\varepsilon _{\mathrm{G}}$ at
various $\mu _{\mathrm{D,\uparrow }}$, which measures $K_{I}=e\Gamma \left(
\mu _{\mathrm{D,\uparrow }}-\mu _{\mathrm{S,\uparrow }}\right) $. Therefore, 
$\Gamma $ and $\mu _{\mathrm{S,\uparrow }}$ are obtained from the linear fit
to the plot of $K_{I}\left( \mu _{\mathrm{D,\uparrow }}\right) $. The error $%
\delta V_{\mathrm{eff}}$ in the estimate of $V_{\mathrm{eff}}=\left( \mu _{%
\mathrm{D,\uparrow }}-\mu _{\mathrm{S,\uparrow }}\right) /e$ is obtained
from the residuals of the fit.

Second, we evaluate energy current $K_{J,\varepsilon _{\mathrm{G}}}=\frac{-1%
}{eh\Gamma }\int \varepsilon _{\mathrm{G}}I_{\mathrm{D}}d\varepsilon _{%
\mathrm{G}}$ at various $\mu _{\mathrm{D,\uparrow }}$, which measures $%
K_{J,\varepsilon _{\mathrm{G}}}=\Delta J-\frac{1}{eh\Gamma }\left( \bar{\mu}-%
\tilde{\varepsilon}\right) K_{I}$. Here, $\Delta J=J_{\mathrm{S,\uparrow }%
}-J_{\mathrm{D,\uparrow }}$ is the difference of the heat current $J_{r}=%
\frac{1}{h}\int \left( E-\mu _{r}\right) \left[ f_{r}\left( E-\mu
_{r}\right) -\theta \left( \mu _{r}-E\right) \right] dE$ between the source (%
$r=$ S$\mathrm{,\uparrow }$) and the drain ($r=$ D$\mathrm{,\uparrow }$). $%
\Delta J$ can be estimated from the $K_{J,\varepsilon _{\mathrm{G}}}$ value
for the $\varepsilon _{\mathrm{G}}$ scan at $V_{\mathrm{eff}}=0$ ($K_{I}=0$%
). Therefore, $\tilde{\varepsilon}-\bar{\mu}=eh\Gamma \left( K_{J}-\Delta
J\right) /K_{I}$ can be estimated for other scans at $V_{\mathrm{eff}}\neq 0$
($K_{I}\neq 0$).

In principle, charge fluctuations as well as the level shifts in $\tilde{%
\varepsilon}$ can be included to obtain $\varepsilon =$ $\varepsilon _{%
\mathrm{G}}+\tilde{\varepsilon}$. However, this correction does not work
precisely at small $\left\vert V_{\mathrm{eff}}\right\vert $ where both $%
\left( K_{J,\varepsilon _{\mathrm{G}}}-\Delta J\right) $ and $K_{I}$ are
small quantities as compared to the experimental noise. Therefore, we
correct large and rare switching events seen as jumps in $K_{J,\varepsilon _{%
\mathrm{G}}}$ at large $\left\vert V_{\mathrm{eff}}\right\vert $ by using
the above formula, and other fluctuations in $\tilde{\varepsilon}$ are
ignored by the linear fit to $\tilde{\varepsilon}\left( \mu _{\mathrm{%
D,\uparrow }}\right) $ in between the switching events. The error $\delta
\varepsilon $ in the estimate of $\varepsilon $ is obtained from the
residuals of the fit at large $\left\vert V_{\mathrm{eff}}\right\vert $.

The above conversion scheme works well for our data. Here, we present a
representative data conversion for Fig. 4a. Because we know finite $\mu _{%
\mathrm{S,\uparrow }}$ appears in our setup, the drain chemical potential $%
\mu _{\mathrm{D,\uparrow }}=\mu _{\mathrm{D,\uparrow }}^{\prime }+\mu _{%
\mathrm{D,0}}$ is varied by $\mu _{\mathrm{D,\uparrow }}^{\prime }$ ($=eV_{%
\mathrm{D}}^{\prime }$) from a temporary chosen offset $\mu _{\mathrm{D,0}}$
($=eV_{\mathrm{D,0}}$) near the expected $\mu _{\mathrm{S,\uparrow }}$. The
original data $I_{\mathrm{D}}\left( V_{\mathrm{GD}},\mu _{\mathrm{D,\uparrow 
}}^{\prime }\right) $ obtained by sweeping $V_{\mathrm{GD}}$ at various $\mu
_{\mathrm{D,\uparrow }}^{\prime }$ is shown in Fig. 4a, where the highly
conductive regions (pink for $I_{\mathrm{D}}>$ 5 pA and cyan for $I_{\mathrm{%
D}}<$ -5 pA) roughly determine the transport window ($\mu _{\mathrm{%
S,\uparrow }}<\varepsilon <\mu _{\mathrm{D,\uparrow }}$ and $\mu _{\mathrm{%
D,\uparrow }}<\varepsilon <\mu _{\mathrm{S,\uparrow }}$). Taking a closer
look at the boundaries of the highly conductive regions, one can see an
initial drift (marked by X) induced by the previous QD condition just before
starting this measurement from $\mu _{\mathrm{D,\uparrow }}^{\prime }=$ 150 $%
\mu $eV, a large jump (marked by Y) associated with switching of nearby
impurity states, and the continuous electrostatic shift (not visible as it
is tiny) due to the variation of $\mu _{\mathrm{D,\uparrow }}^{\prime }$.
Since positive and negative $I_{\mathrm{D}}$ regions show a complicated
pattern, we should rely on the conversion scheme to determine the conditions
for $V_{\mathrm{eff}}=0$ and $\varepsilon =\bar{\mu}$.

The integrated current $K_{I}=\int I_{\mathrm{D}}d\varepsilon _{\mathrm{G}}$
is numerically obtained with $\alpha =$ 0.036, as shown in Fig. 4b. $K_{I}$
increases almost linearly with $\mu _{\mathrm{D}}^{\prime }$, which can be
fitted with the form $K_{I}=e\Gamma \left( \mu _{\mathrm{D,\uparrow }%
}^{\prime }-\mu _{\mathrm{S,\uparrow }}^{\prime }\right) $ by defining $\mu
_{\mathrm{S,\uparrow }}^{\prime }$ measured from the temporal offset ($\mu _{%
\mathrm{S,\uparrow }}=\mu _{\mathrm{S,\uparrow }}^{\prime }+\mu _{\mathrm{D,0%
}}$), as shown by the red line. The fitting suggests $e\Gamma =$ 20 pA and $%
\mu _{\mathrm{S,\uparrow }}^{\prime }=$ -2.2 $\mu $eV, from which $eV_{%
\mathrm{eff}}=\mu _{\mathrm{D,\uparrow }}^{\prime }-\mu _{\mathrm{S,\uparrow 
}}^{\prime }$ is obtained for this data set. The data points scattered
around the fit imply the error $\delta V_{\mathrm{eff}}\simeq $ 1 $\mu $eV
(see the inset at around $\mu _{\mathrm{D,\uparrow }}^{\prime }=0$).

Next, the energy current $K_{J,\varepsilon _{\mathrm{G}}}=\frac{-1}{eh\Gamma 
}\int \varepsilon _{\mathrm{G}}I_{\mathrm{D}}d\varepsilon _{\mathrm{G}}$ is
numerically obtained, as shown in Fig. 4c. The $K_{J,\varepsilon _{\mathrm{G}%
}}$ value at $\mu _{\mathrm{D,\uparrow }}^{\prime }=\mu _{\mathrm{S,\uparrow 
}}^{\prime }$ (the vertical dashed line) provides $\Delta J=$ 190 fW. The
initial drift and the large jump are detected in the $\mu _{\mathrm{%
D,\uparrow }}^{\prime }$ dependence of $K_{J,\varepsilon _{\mathrm{G}}}$, as
marked by X and Y, respectively. Then, $\tilde{\varepsilon}-\bar{\mu}%
=eh\Gamma \left( K_{J,\varepsilon _{\mathrm{G}}}-\Delta J\right) /K_{I}$ is
obtained, as shown by the black line in Fig. 4d. The large fluctuation
around $\mu _{\mathrm{D,\uparrow }}^{\prime }=\mu _{\mathrm{S,\uparrow }%
}^{\prime }$ comes from finite noise in the small numerator $\left(
K_{J,\varepsilon _{\mathrm{G}}}-\Delta J\right) $ and denominator $K_{I}$,
where $\left( \tilde{\varepsilon}-\bar{\mu}\right) $ values are unreliable.
Otherwise, the $\mu _{\mathrm{D,\uparrow }}^{\prime }$ dependence shows the
initial drift (marked by X), the large jump (marked by Y), and the
continuous electrostatic shift with the averaged slope of $\beta =d\left( 
\tilde{\varepsilon}-\bar{\mu}\right) /d\mu _{\mathrm{D,\uparrow }}^{\prime
}\simeq $ -0.15 due to the electrostatic shift from varying $\mu _{\mathrm{%
D,\uparrow }}^{\prime }$. This $\beta $ value is consistent with the
asymmetry of the Coulomb diamond obtained for a wide range of $V_{\mathrm{eff%
}}$ in the absence of heat injection. For this data, jumps greater than 12
fW in $K_{J,\varepsilon _{\mathrm{G}}}$ [marked by the vertical bars in Fig.
4d] are corrected by the above formula, while other fluctuations are
ignored, as shown by the red line. We use this $\tilde{\varepsilon}-\bar{\mu}
$ (the red line) to obtain $\varepsilon -\bar{\mu}$. The ignored
fluctuations at large $\left\vert \mu _{\mathrm{D,\uparrow }}^{\prime }-\mu
_{\mathrm{S,\uparrow }}^{\prime }\right\vert $ are considered as the error $%
\delta \varepsilon \simeq $ 5 $\mu $eV in the estimate of $\varepsilon $.

With $eV_{\mathrm{eff}}$ and $\varepsilon -\bar{\mu}$ obtained by the above
scheme, $I_{\mathrm{D}}$ is replotted as a function of $eV_{\mathrm{eff}}$
and $\varepsilon -\bar{\mu}$ in Fig. 4e. The initial drift and the large
jump are removed in $I_{\mathrm{D}}\left( \varepsilon -\bar{\mu},eV_{\mathrm{%
eff}}\right) $ plot. The boundary between the positive and negative current
regions is now smoothly connected. The boundary is not linear, i.e., its
slope $d\left( eV_{\mathrm{eff}}\right) /d\left( \varepsilon -\bar{\mu}%
\right) $ is gentle at $\varepsilon -\bar{\mu}\simeq 0$ and steep at large $%
\left\vert \varepsilon -\bar{\mu}\right\vert \gtrsim k_{\mathrm{B}}T_{%
\mathrm{B}}$. The scheme determines the conditions for $eV_{\mathrm{eff}}=0$
and $\varepsilon -\bar{\mu}=0$, from which we draw the idealized efficiency $%
\bar{\eta}=eV_{\mathrm{eff}}/\left( \varepsilon -\mu _{\mathrm{S,\uparrow }%
}\right) $ for $\bar{\eta}=$ 0.2, 0.4, and 0.6 in Fig. 2c. The same
conversion scheme is applied to all data obtained for NT and QT states in
setups I and II. The error bars in $\bar{\eta}$ of Figs. 6b and 6d show the
range from $e\left( V_{\mathrm{eff}}-\delta V_{\mathrm{eff}}\right) /\left(
\varepsilon +\delta \varepsilon -\mu _{\mathrm{S,\uparrow }}\right) $ to $%
e\left( V_{\mathrm{eff}}+\delta V_{\mathrm{eff}}\right) /\left( \varepsilon
-\delta \varepsilon -\mu _{\mathrm{S,\uparrow }}\right) $ for each data
point by considering the errors $\delta \varepsilon $ and $\delta V_{\mathrm{%
eff}}$.

\bigskip

\textbf{Energy-harvesting performances}. The performances of the energy
harvesting in Fig. 6 are evaluated in the following way. We show the
representative analysis in Supplementary Fig. 4 by using the converted data
in Fig. 2a (the NT state at $g=$ 0.028 and $J_{\mathrm{T}}=$ 380 fW) and
Fig. 2b (the QT state at $g=$ 0.45 and $J_{\mathrm{T}}=$ 420 fW) taken in
setup I.

Maximum power $P_{\mathrm{M}}$: For each data [$I_{\mathrm{D}}\left(
\varepsilon -\bar{\mu},eV_{\mathrm{eff}}\right) $ as a function of $%
\varepsilon -\bar{\mu}$ at various $eV_{\mathrm{eff}}$], we record the
constrained minimum $I_{\mathrm{D,\min }}$ ($<0$) among all $\varepsilon -%
\bar{\mu}$ values at each $eV_{\mathrm{eff}}$ and plot the constrained
maximum power $P_{\mathrm{m}}=-I_{\mathrm{D,\min }}V_{\mathrm{eff}}$ as a
function of $eV_{\mathrm{eff}}$ in Supplementary Fig. 4a. Here, $P_{\mathrm{m%
}}$ is normalized by $\Gamma $, because $P_{\mathrm{m}}$ should change
linearly with $\Gamma $ which is different for the NT ($e\Gamma \simeq $ 20
pA) and QT data ($\simeq $ 56 pA). The length of each bar shows the error
associated with the current noise $\delta I_{\mathrm{D}}\simeq $ 50 fA and $%
\delta V_{\mathrm{eff}}\simeq $ 1 $\mu $eV. The maximum power generation $P_{%
\mathrm{M}}/\Gamma $ (marked by `max') and the electromotive force $V_{%
\mathrm{emf}}$ (the arrows) are determined from this plot for each data set.
This $V_{\mathrm{emf}}$ is plotted in Fig. 6a, and $R=P_{\mathrm{M}}/J_{%
\mathrm{T}}\Gamma $ is plotted in Fig. 6c.

Efficiency $\bar{\eta}_{\mathrm{M}}$ at maximum power: The condition $%
\varepsilon _{\mathrm{m}}-\bar{\mu}$ at the constrained minimum $I_{\mathrm{%
D,\min }}$ is also recorded to plot the corresponding efficiency $\bar{\eta}%
_{\mathrm{m}}=eV_{\mathrm{eff}}/\left( \varepsilon _{\mathrm{m}}-\mu _{%
\mathrm{S}}\right) $ in Supplementary Fig. 4b. The length of each bar shows
the error associated with $\delta V_{\mathrm{eff}}\simeq $ 1 $\mu $eV and $%
\delta \varepsilon \simeq $ 5 $\mu $eV. The $\bar{\eta}_{\mathrm{m}}$ value
at the maximum power generation $P_{\mathrm{M}}$ (connected by the vertical
dashed lines) represents the efficiency $\bar{\eta}_{\mathrm{M}}$ at maximum
power. This $\bar{\eta}_{\mathrm{M}}$ is plotted in Fig. 6d.

The maximum efficiency $\bar{\eta}_{\mathrm{Z}}$ in the zero power limit:
The condition $\varepsilon _{\mathrm{z}}-\bar{\mu}$ at the boundary between
positive and negative $I_{\mathrm{D}}$ regions is recorded at each $eV_{%
\mathrm{eff}}$, and the corresponding efficiency $\bar{\eta}_{\mathrm{z}%
}=eV_{\mathrm{eff}}/\left( \varepsilon _{\mathrm{z}}-\mu _{\mathrm{S}%
}\right) $ is plotted in Supplementary Fig. 4c. The length of each bar shows
the error associated with $\delta V_{\mathrm{eff}}\simeq $ 1 $\mu $eV and $%
\delta \varepsilon \simeq $ 5 $\mu $eV. $\bar{\eta}_{\mathrm{z}}$ increases
with increasing $eV_{\mathrm{eff}}$ for the NT data but decreases for the QT
data. The maximum efficiency $\bar{\eta}_{\mathrm{Z}}$ in the zero power
limit is determined from the maximum $\bar{\eta}_{\mathrm{z}}$ in this plot.
The $\bar{\eta}_{\mathrm{Z}}$ value for the QT state involves large
uncertainty, as it appears at small $eV_{\mathrm{eff}}$. Nevertheless, the $%
\bar{\eta}_{\mathrm{Z}}$ value for the NT state is greater than the averaged 
$\bar{\eta}_{\mathrm{Z}}$ value for the QT state. This $\bar{\eta}_{\mathrm{Z%
}}$ is plotted in Fig. 6b.

The above scheme is applied to the simulated data $I_{\mathrm{D}}\left(
\varepsilon ,eV_{\mathrm{eff}}\right) $ for the NT and thermalized states
under the same $J_{\mathrm{T}}=$ 420 fW, as shown in Supplementary Figs.
4d-f. The calculated $P_{\mathrm{m}}/\Gamma $, $\bar{\eta}_{\mathrm{m}}$,
and $\bar{\eta}_{\mathrm{z}}$ reproduce the experimental features quite
well. Particularly, large $\bar{\eta}_{\mathrm{M}}$ for the NT state exceeds
the Curzon--Ahlborn efficiency $\eta _{\mathrm{CA}}$ in Supplementary Fig.
4e, and large $\bar{\eta}_{\mathrm{Z}}$ for the NT state exceeds the Carnot
efficiency $\eta _{\mathrm{C}}$ in Supplementary Fig. 4f. The simulated $V_{%
\mathrm{emf}}$, $\bar{\eta}_{\mathrm{Z}}$, $\bar{\eta}_{\mathrm{M}}$, $\eta
_{\mathrm{C}}$, and $\eta _{\mathrm{CA}}$ are shown by solid lines in Fig.
6a-d.

\bigskip

\textbf{Multi-level simulation.} We employed the standard master equation
for occupation probability $p_{N,n}$ of $n$-th level ($n=0$ for the ground
state and $n\geq 1$ for the $n$-th excited states) of $N$-electron QD \cite%
{Itoh-PRL2018}. Tunneling transition between $n$-th level $\Delta _{n}$ of $%
N $-electron QD and $m$-th level $\Delta _{m}^{\prime }$ of $\left(
N+1\right) $-electron QD, where the energy levels $\Delta _{n}$ and $\Delta
_{m}^{\prime }$ are measured from the respective ground state, is
characterized by the corresponding chemical potential $\mu _{nm}=\varepsilon
+\Delta _{m}^{\prime }-\Delta _{n}$ and tunneling rates $\Gamma
_{nm}^{\left( \mathrm{S}\right) }$ on the source side and $\Gamma
_{nm}^{\left( \mathrm{D}\right) }$ on the drain side. The lifetime
broadening is neglected for simplicity.\ By solving the steady state
condition ($\frac{d}{dt}p_{N,n}=0$) of the master equation, the current $I_{%
\mathrm{D}}$ is calculated as a function of $\varepsilon $ and $eV_{\mathrm{%
eff}}$. In the same manner as the single-level model, we extract $P_{\mathrm{%
M}}/J_{\mathrm{T}}\Gamma $ from the simulated data $I_{\mathrm{D}}\left(
\varepsilon ,eV_{\mathrm{eff}}\right) $ and plot it as a function of $J_{%
\mathrm{T}}$, as shown by the dashed lines in Fig. 6c. Here, we used the
parameters $\Delta _{1}=$ 400 $\mu $eV, $\Delta _{1}^{\prime }=$ 80 $\mu $%
eV, $\Delta _{2}^{\prime }=$ 400 $\mu $eV, $\Gamma _{nm}^{\left( \mathrm{S}%
\right) }/\Gamma _{nm}^{\left( \mathrm{D}\right) }=$ 0.1, $\Gamma
_{10}^{\left( \mathrm{S/D}\right) }/\Gamma _{00}^{\left( \mathrm{S/D}\right)
}=$ 6, $\Gamma _{01}^{\left( \mathrm{S/D}\right) }/\Gamma _{00}^{\left( 
\mathrm{S/D}\right) }=$ 1, and $\Gamma _{02}^{\left( \mathrm{S/D}\right)
}/\Gamma _{00}^{\left( \mathrm{S/D}\right) }=$ 10. Compared to the
simulation with the single-level model, $P_{\mathrm{M}}/J_{\mathrm{T}}\Gamma 
$ becomes larger particularly for the NT state, because high-energy
electrons can pass through the excited states.\bigskip

{\large Data availability}

The data and analysis used in this work are available in the main text, 
Supplementary Figures, and Supplementary Data 1. Any other relevant data 
are available from the corresponding author upon reasonable request.

\bigskip

\textbf{Acknowledgements }We thank Shunya Akiyama and Ryota Konuma for
preliminary measurements in the early stage of the study and Haruki Minami,
Yasuhiro Tokura, and Keiji Saito for fruitful discussions. This study was
supported by the Grants-in-Aid for Scientific Research (KAKENHI JP19H05603
and JP23K17302) and \textquotedblleft Advanced Research Infrastructure for
Materials and Nanotechnology in Japan (ARIM)\textquotedblright\ program of
the Ministry of Education, Culture, Sports, Science and Technology (MEXT),
Japan.

\bigskip

\textbf{Author contributions }T.F. conceived and supervised the project.
T.A. and K.M. grew the heterostructure, and C.L. fabricated the device. H.Y.
and M.U. performed the measurements with help from H.T., T.H., and T.F.
H.Y., M.U., and T.F. analyzed the data and wrote the paper. All authors
discussed the results.

\bigskip

\textbf{Competing interests }The authors declare no competing interests.


\begin{thebibliography}{99}
\bibitem{MyersAVS2022} N. M. Myers, O. Abah, and S. Deffner, Quantum
thermodynamic devices: From theoretical proposals to experimental reality,
AVS Quantum Science \textbf{4}, 027101 (2022).

\bibitem{BookQuantumThermodynamics} S. Deffner and S. Campbell, Quantum
Thermodynamics (Morgan \& Claypool, San Rafael, 2019).

\bibitem{ScullyScience2003} M. O. Scully, M. S. Zubairy, G. S. Agarwal, and
H. Walther, Extracting Work from a Single Heat Bath via Vanishing Quantum
Coherence, Science \textbf{299}, 862 (2003).

\bibitem{RossnagelPRL2014} J. Ro\ss nagel, O. Abah, F. Schmidt-Kaler, K.
Singer, and E. Lutz, Nanoscale Heat Engine Beyond the Carnot Limit, Phys.
Rev. Lett. \textbf{112}, 030602 (2014).

\bibitem{KlaersPRX2017} J. Klaers, S. Faelt, A. Imamoglu, and E. Togan,
Squeezed Thermal Reservoirs as a Resource for a Nanomechanical Engine beyond
the Carnot Limit, Phy. Rev. X \textbf{7}, 031044 (2017).

\bibitem{KimNatPhoto2022} J. Kim, S.-h. Oh, D. Yang, J. Kim, M. Lee, and K.
An, A photonic quantum engine driven by superradiance, Nat. Photonics 
\textbf{16}, 707 (2022).

\bibitem{MayerCommPhys2023} D. Mayer, E. Lutz, and A. Widera, Generalized
Clausius inequalities in a nonequilibrium cold-atom system, Communications
Physics \textbf{6}, 61 (2023).

\bibitem{KennesPRB2017} D. M. Kennes, Phys. Rev. B \textbf{96}, 024302
(2017).

\bibitem{ChenNPJQI2019} Y.-Y. Chen, G. Watanabe, Y.-C. Yu, X.-W. Guan and A.
del Campo, npj Quantum Information \textbf{5}, 88 (2019).

\bibitem{KinoshitaNature2006} T. Kinoshita, T. Wenger, and D. S. Weiss, A
quantum Newton's cradle, Nature \textbf{440}, 900 (2006).

\bibitem{BlochRMP2008} I. Bloch, J. Dalibard, and W. Zwerger, Many-body
physics with ultracold gases, Rev. Mod. Phys. \textbf{80}, 885 (2008).

\bibitem{KollarPRB2011} M. Kollar, F. A. Wolf, and M. Eckstein, Generalized
Gibbs ensemble prediction of prethermalization plateaus and their relation
to nonthermal steady states in integrable systems, Phys. Rev. B \textbf{84},
054304 (2011).

\bibitem{BookGiamarchi} T. Giamarchi, Quantum Physics in One Dimension
(Oxford Univ. Press, 2004).

\bibitem{WashioPRB2016} K. Washio, R. Nakazawa, M. Hashisaka, K. Muraki, Y.
Tokura, and T. Fujisawa, Long-lived binary tunneling spectrum in the quantum
Hall Tomonaga-Luttinger liquid, Phys. Rev. B \textbf{93}, 075304 (2016).

\bibitem{Itoh-PRL2018} K. Itoh, R. Nakazawa, T. Ota, M. Hashisaka, K. Muraki
and T. Fujisawa, Signatures of a Nonthermal Metastable State in
Copropagating Quantum Hall Edge Channels. Phys. Rev. Lett. \textbf{120},
197701 (2018).

\bibitem{Altimiras-NatPhys10} C. Altimiras, H. le Sueur, U. Gennser, A.
Cavanna, D. Mailly, and F. Pierre, Non-equilibrium edge-channel spectroscopy
in the integer quantum Hall regime. Nat. Phys. \textbf{6}, 34 (2010).

\bibitem{leSueurPRL2010} H. le Sueur, C. Altimiras, U. Gennser, A. Cavanna,
D. Mailly, and F. Pierre, Energy Relaxation in the Integer Quantum Hall
Regime, Phys. Rev. Lett. \textbf{105}, 056803 (2010).

\bibitem{JezouinScience2013} S. Jezouin, F. D. Parmentier, A. Anthore, U.
Gennser, A. Cavanna, Y. Jin and F. Pierre, Quantum Limit of Heat Flow Across
a Single Electronic Channel, Science \textbf{342}, 601 (2013).

\bibitem{SivreNatComm2019} E. Sivre, H. Duprez, A. Anthore, A. Aassime, F.
D. Parmentier, A. Cavanna, A. Ouerghi, U. Gennser, and F. Pierre, Electronic
heat flow and thermal shot noise in quantum circuits, Nat. Commun. \textbf{10%
}, 5638 (2019).

\bibitem{KonumaPRB2022} R. Konuma, C. Lin, T. Hata, T. Hirasawa, T. Akiho,
K. Muraki, and T. Fujisawa, Nonuniform heat redistribution among multiple
channels in the integer quantum Hall regime, Phys. Rev. B \textbf{105},
235302 (2022).

\bibitem{SivreNatPhys2018} E. Sivre, A. Anthore, F. D. Parmentier, A.
Cavanna, U. Gennser, A. Ouerghi, Y. Jin and F. Pierre, Heat Coulomb blockade
of one ballistic channel, Nat. Phys. 14, 145 (2018).

\bibitem{RosenblattNatComm2017} A. Rosenblatt, F. Lafont, I. Levkivskyi, R.
Sabo, I. Gurman, D. Banitt, M. Heiblum, and V. Umansky, Transmission of heat
modes across a potential barrier, Nat. Commun. 8, 2251 (2017).

\bibitem{RouraBasPRB2018} P. Roura-Bas, L. Arrachea and E. Fradkin, Enhanced
thermoelectric response in the fractional quantum Hall effect, Phys. Rev. B
97, 081104 (2018).

\bibitem{FreulonNatComm-SC} V. Freulon, A. Marguerite, J. M. Berroir, B. Pla%
\c{c}ais, A. Cavanna, Y. Jin and G. F\'{e}ve, Hong-Ou-Mandel experiment for
temporal investigation of single-electron fractionalization, Nat. Commun. 
\textbf{6}, 6854 (2015).

\bibitem{Bocquillon-NatCom2013} E. Bocquillon et al., Separation of neutral
and charge modes in one-dimensional chiral edge channels. Nat. Commun. 
\textbf{4}, 1839 (2013).

\bibitem{Inoue-PRL2014} H. Inoue, A. Grivnin, N. Ofek, I. Neder, M. Heiblum,
V. Umansky, and D. Mahalu, Charge fractionalization in the integer quantum
Hall effect, Phys. Rev. Lett. \textbf{112}, 166801 (2014).

\bibitem{Hashisaka-NatPhys2017} M. Hashisaka, N. Hiyama, T. Akiho, K.
Muraki, and T. Fujisawa, Waveform measurement of charge- and spin-density
wavepackets in a chiral Tomonaga--Luttinger liquid, Nat. Phys. \textbf{13},
559 (2017).

\bibitem{GutmanPRL2008} D. B. Gutman, Y. Gefen, and A. D. Mirlin,
Nonequilibrium Luttinger Liquid: Zero-Bias Anomaly and Dephasing, Phys. Rev.
Lett. \textbf{101}, 126802 (2008).

\bibitem{Iucci2009} A. Iucci and M. A. Cazalilla, Quantum quench dynamics of
the Luttinger model, Phys. Rev. A \textbf{80}, 063619 (2009).

\bibitem{GutmanPRB2010} D. B. Gutman, Y. Gefen, and A. D. Mirlin,
Bosonization of one-dimensional fermions out of equilibrium, Phys. Rev. B 
\textbf{81}, 085436 (2010).

\bibitem{KovrizhinPRB2011} D. L. Kovrizhin and J. T. Chalker, Equilibration
of integer quantum Hall edge states, Phys. Rev. B \textbf{84}, 085105 (2011).

\bibitem{LevkivskyiPRB2012} I. P. Levkivskyi and E. V. Sukhorukov, Energy
relaxation at quantum Hall edge, Phys. Rev. B \textbf{85}, 075309 (2012).

\bibitem{SanchezPRL2019} R. S\'{a}nchez, J. Splettstoesser, and R. S.
Whitney, Nonequilibrium System as a Demon, Phys. Rev. Lett. 123, 216801
(2019).

\bibitem{EspositoEuroPhysLett2009} M. Esposito, K. Lindenberg, and C. Van
den Broeck, Thermoelectric efficiency at maximum power in a quantum dot.
Europhys. Lett. 85, 60010 (2009).

\bibitem{JosefssonNatNano2018} M. Josefsson, A. Svilans, A. M. Burke, E. A.
Hoffmann, S. Fahlvik, C. Thelander, M. Leijnse, and H. Linke, A quantum-dot
heat engine operating close to the thermodynamic efficiency limits, Nat.
Nanotechnol. \textbf{13}, 920 (2018).

\bibitem{HumphreyPRL2002} T. E. Humphrey, R. Newbury, R. P. Taylor, and H.
Linke, Reversible Quantum Brownian Heat Engines for Electrons, Phys. Rev.
Lett. 89, 116801 (2002).

\bibitem{SuzukiCommPhys2023} K. Suzuki, T. Hata, Y. Sato, T. Akiho, K.
Muraki, and T. Fujisawa, Non-thermal Tomonaga-Luttinger liquid eventually
emerging from hot electrons in the quantum Hall regime, Communications
Physics \textbf{6}, 103 (2023).

\bibitem{PothierPRL1997} H. Pothier, S. Gu\'{e}ron, N. O. Birge, D. Esteve
and M. H. Devoret, Energy Distribution Function of Quasiparticles in
Mesoscopic Wires, Phys. Rev. Lett. \textbf{79}, 3490-3493 (1997).

\bibitem{ProkudinaPRL2014} M. G. Prokudina, S. Ludwig, V. Pellegrini, L.
Sorba, G. Biasiol and V. S. Khrapai, Tunable Nonequilibrium Luttinger Liquid
Based on Counterpropagating Edge Channels, Phys. Rev. Lett. \textbf{112},
216402 (2014).

\bibitem{vonDelftAnnPhys1998} J. von Delft and H. Schoeller, Bosonization
for beginners --- refermionization for experts, \textit{Ann. Phys.} \textbf{7%
}, 225 (1998).

\bibitem{FujisawaAnnPhys2022} T. Fujisawa, Nonequilibrium Charge Dynamics of
Tomonaga--Luttinger Liquids in Quantum Hall Edge Channels, Ann. Phys.
(Berlin) \textbf{534,} 2100354 (2022).

\bibitem{RodriguezNatComm2020} R. H. Rodriguez, F. D. Parmentier, D.
Ferraro, P. Roulleau, U. Gennser, A. Cavanna, M. Sassetti, F. Portier, D.
Mailly and P. Roche, Relaxation and revival of quasiparticles injected in an
interacting quantum Hall liquid, Nat. Commun. \textbf{11}, 2426 (2020).

\bibitem{HashisakaPRB2012} M. Hashisaka, K. Washio, H. Kamata, K. Muraki,
and T. Fujisawa, Distributed electrochemical capacitance evidenced in
high-frequency admittance measurements on a quantum Hall device, Phys. Rev.
B \textbf{85}, 155424 (2012).

\bibitem{WhitneyPRL2014} R. S. Whitney, Most Efficient Quantum
Thermoelectric at Finite Power Output, Phys. Rev. Lett. \textbf{112}, 130601
(2014).

\bibitem{TaruchaSSC1995} S. Tarucha, T. Honda, and T. Saku, Reduction of
quantized conductance at low temperatures observed in 2 to 10 $\mu $m-long
quantum wires, Solid State Commun. \textbf{94}, 413 (1995).

\bibitem{BockrathNature1999} M. Bockrath, D. H. Cobden, J. Lu, A. G.
Rinzler, R. E. Smalley, L. Balents, and P. L. McEuen, Luttinger-liquid
behaviour in carbon nanotubes, Nature \textbf{397}, 598 (1999).

\bibitem{ChangRMB} A. M. Chang, Chiral Luttinger liquids at the fractional
quantum Hall edge, Rev. Mod. Phys. \textbf{75}, 1449 (2003).

\bibitem{AuslaenderScience2005} O. M. Auslaender, H. Steinberg, A. Yacoby,
Y. Tserkovnyak, B. I. Halperin, K. W. Baldwin, L. N. Pfeiffer, and K. W.
West, Spin-Charge Separation and Localization in One Dimension, Science 
\textbf{308}, 88 (2005).

\bibitem{UedaNRP2020} M. Ueda, Quantum equilibration, thermalization and
prethermalization in ultracold atoms, Nat. Rev. Physics \textbf{2}, 669
(2020).
\end{thebibliography}
\end{document}